\documentclass[journal,final, onecolumn]{IEEEtran}

\usepackage{graphicx}

\usepackage{cite}

\ifCLASSINFOpdf

\else

\fi

\usepackage[cmex10]{amsmath}
\usepackage{amssymb}
\usepackage{subeqnarray}
\DeclareMathOperator{\tr}{tr}
\DeclareMathOperator{\cov}{cov}

\allowdisplaybreaks

\begin{document}

\sloppy
\title{Vector Gaussian Successive Refinement With Degraded Side Information}


\author{Yinfei~Xu,~\IEEEmembership{Member,~IEEE,}
  Xuan~Guang,~\IEEEmembership{Member,~IEEE,}
  Jian~Lu,~\IEEEmembership{Member,~IEEE,} \\and Jun~Chen,~\IEEEmembership{Senior Member,~IEEE}
\thanks{Y. Xu and J. Lu are with the School of Information Science and Engineering, Southeast University, Nanjing, 210096, China (email:  \{yinfeixu, lujian1980\}@seu.edu.cn).}
\thanks{X. Guang is with the School of Mathematical Sciences and LPMC, Nankai University, Tianjin, 300071, China (email: xguang@nankai.edu.cn).}
\thanks{J. Chen is with the Department of Electrical and Computer Engineering, McMaster University, ON, L8S 4K1, Canada (email: junchen@ece.mcmaster.ca).}
\thanks{This paper was presented in part at the 2019 IEEE International Symposium on Information Theory.}
}

\maketitle

\begin{abstract}
%\boldmath
We investigate the problem of the successive refinement for Wyner-Ziv coding with degraded side information and obtain a complete characterization of the rate region for the quadratic vector Gaussian case. The achievability part is based on the evaluation of the Tian-Diggavi inner bound that involves Gaussian auxiliary random vectors. For the converse part, a matching outer bound is obtained with the aid of a new extremal inequality. Herein, the proof of this extremal inequality depends on the integration of the monotone path argument and the doubling trick as well as information-estimation relations.

%In this paper, we consider Successive refinement for the Wyner-Ziv source coding problem, in which sources are vector Gaussian distributed and side information follows a degraded order assumption. We fully characterize its rate-distortion region with covariance mean square error distortions. We use jointly Gaussian auxiliary random variables to evaluate the existing single-letter description of the rate region by Tian and Diggavi, which gives the achievability. In the converse part, a new extremal inequality is established, which is further leverage to find a tight outer bound of the rate-distortion region. The proof of the extremal inequality is based on the recent discovered Gaussian perturbation construction, and combined with information-estimation relationship, suitably.
\end{abstract}

\begin{IEEEkeywords}
Extremal inequality, lossy source coding, mean squared error, rate region, side information, successive refinement, vector Gaussian source, Wyner-Ziv problem.
\end{IEEEkeywords}

\newtheorem{theorem}{Theorem}
\newtheorem{lemma}{Lemma}
\newtheorem{definition}{Definition}
\newtheorem{remark}{Remark}
\newtheorem{example}{Example}

% \fi

\section{Introduction}
The research on network source coding can be traced back to the seminal work by Slepian and Wolf \cite{SW73}, where they considered, among other things, the problem of lossless source coding with side information at the decoder. Wyner and Ziv \cite{WZ76} studied the lossy source coding version of this problem (which later bears their names) and characterized its information-theoretic limit. Subsequently, the Wyner-Ziv problem was extended in various ways. One particular extension, known as successive refinement for Wyner-Ziv coding with degraded side information, is as follows: A source is encoded and decoded, in a successive manner, to meet different distortion constraints with the aid of progressively enhanced decoder side information. This extended Wyner-Ziv problem was tackled by Steinberg and Merhav \cite{SM04} for the two-stage case and by Tian and Diggavi \cite{TD07} for the multi-stage case. Specifically, the computable characterizations of rate regions in the discrete memoryless setting (with a general distortion measure) and in the scalar Gaussian setting (with the quadratic error distortion measure) were obtained accordingly.

In this paper, we consider the same extended Wyner-Ziv problem with a particular attention paid to the vector Gaussian setting (under covariance distortion constraints). The heart of the present paper is a new inequality regarding the optimality of the Gaussian solution to a certain extremal problem. It is well known that extremal inequalities play an important role in characterizing the fundamental limits of Gaussian network source and channel coding problems is well known. Indeed, they are indispensable to the converse argument for the Gaussian broadcast channel coding problem \cite{Bergmans74, WLSSV09,LLL09,EU12-1,EU12-2,LLPS13,EU13,CL14-1,CL14-2,KL14}, the Gaussian interference channel coding problem \cite{MK09, SKC09, AV09}, the Gaussian multi-terminal source coding problem \cite{Oohama05, WCW10, XW13, WC13, WC14}, the secret key generation problem \cite{WO11}, the Gaussian multiple description problem \cite{Ozarow80, WV07,C09,XCW17}, and others \cite{SCWL13, SCT15}. Basic extremal inequalities that rely on the differential-entropy-maximizing property of the Gaussian distribution can only handle simple situations where the objective functional can be greedily optimized. When there are two or more conflicting terms, Shannon's entropy power inequality is often used to resolve the tension. However, the proportionality condition on the relevant covariance matrices needed for the tightness of the entropy power inequality is quite restrictive, typically only satisfied in scalar source and channel coding problems. As a consequence, more sophisticated extremal inequalities are needed to deal with vector Gaussian sources and channels. The proofs of such extremal inequalities, as well as the proof of the entropy power inequality, are often proved by invoking the monotone path argument or its variants. The conventional monotone path argument nevertheless appears to have its own limitations. For example, it fails to yield a tight outer bound on the capacity region of the two-user vector Gaussian broadcast channel with private and common messages. The desired result is eventually obtained by Geng and Nair \cite{GN14} through a different approach involving so-called doubling trick. On the other hand, this approach obscures some useful information regarding the optimal Gaussian solution. Fortunately, this problem can be remedied via a systematic integration of  the  monotone path argument and the doubling trick, as shown by Wang and Chen \cite{WC19} in their new proof of Courtade's extremal inequality \cite{C18}. In this work, we make use of this integrated strategy, together with the properties of the minimum mean square error (MMSE) and the  Fisher information, to establish a new extremal inequality, which is further leveraged to characterize the rate region of the aforementioned extended Wyner-Ziv problem in the vector Gaussian source setting. It will be seen that the new extremal inequality avoids the comparison of distortion matrices, and thus is particularly handy when dealing with a large number of covariance distortion constraints.

The rest of this paper is organized as follows. We present the problem formulation and the main result in Section \ref{sec:statement}. Section \ref{EI} is devoted to proving a new extremal inequality, which constitutes the main technical part of this paper. The main result is proved in Section~\ref{MT}. We conclude the paper in Section \ref{sec:conclusion}.

%The rest of this paper is organized as follows. The problem setup is given in Section II, we first revisit the rate-distortion region for the case of discrete memoryless sources case which has been characterized in \cite{TD07}, and then we show the rate-distortion characterization for the case of vector Gaussian sources considered in this paper. Section III is devoted to proof our extremal inequality, which is our main technical part of this paper. In Section IV, we compare the supporting hyperplanes of the inner and outer bounds of the rate-distortion region, and establish the achievability and converse of our main result. In Section V, we conclude with a summary of our results and a remark on future research.

%\begin{figure}
%\centering
%\includegraphics[width=0.5\textwidth]{DH}
%\caption{Distributed hypothesis testing from many noisy observations}
%\label{fig}
%\centering
%\end{figure}

\section{Problem Statement and Main Result}\label{sec:statement}

%\subsection{Discrete Memoryless Sources}

\begin{figure}[!t]
\centering
\includegraphics[width=0.8\textwidth]{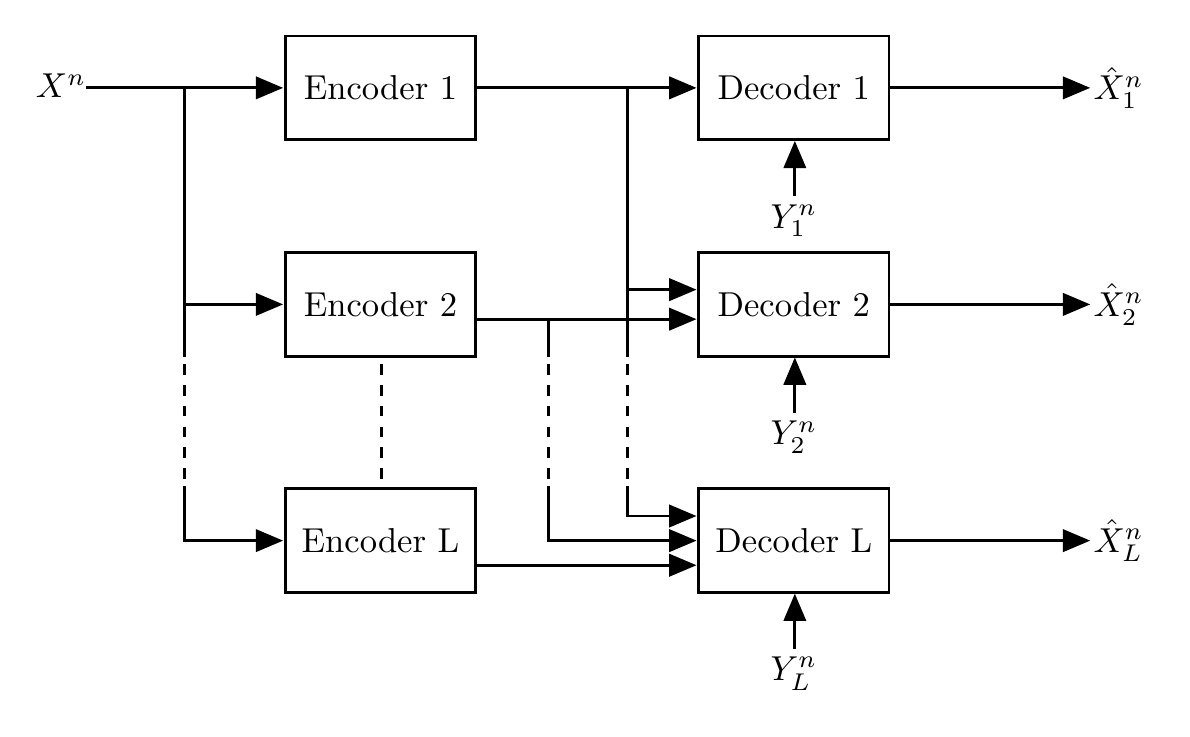}
\caption{Successive refinement for Wyner-Ziv coding with degraded side information.}
\label{fig}
\centering
\end{figure}

Let $X$ be a $p \times 1$-dimensional  random vector with mean zero and covariance matrix $\boldsymbol{K}_{0} \succ \boldsymbol{0}$. Moreover, let
\begin{equation}
{Y}_{i} = {X} + {N}_{i}, \qquad i \in [1:L],
\end{equation}
where ${N}_{i}$ is a $p \times 1$-dimensional random vector with mean zero and covariance matrix $\boldsymbol{K}_{i} \succ \boldsymbol{0}$, $i \in [1:L]$. It is assumed that
\begin{equation}\label{eq:degrad}
\boldsymbol{K}_{1} \succ \ldots \succ \boldsymbol{K}_{L-1} \succ \boldsymbol{K}_{L},
\end{equation}
and $X$, $N_i-N_{i+1}$, $i\in[1:L]$, are mutually independent and jointly Gaussian\footnote{Here $N_{L+1}$ is a null random vector with covariance matrix $\boldsymbol{K}_{L+1}=\mathbf{0}$.}. This assumption implies that
\begin{equation}
X\rightarrow Y_{L} \rightarrow Y_{L-1} \rightarrow \ldots \rightarrow Y_{1}
\end{equation}
forms a Markov chain.  Let $(X(t), Y_i(t), i \in [1:L])_{t=1}^{\infty}$ be i.i.d. copies of  $(X, Y_i, i \in [1:L])$.

The system model can be described as follows (see also Fig. \ref{fig}).
\begin{itemize}
\item $L$ encoding functions $(\phi_{i}^{(n)}, i \in [1:L])$:
\begin{equation}
\phi_{i}^{(n)}: \mathcal{X}^{n} \mapsto \mathcal{M}_{i}^{(n)}, \qquad  i \in [1:L],
\end{equation}
where $\phi_{i}^{(n)}$ maps the source sequence ${X}^{n}$ to the codeword $M_{i}({X}^{n})$, $i \in [1:L]$.

\item $L$ decoding functions $(\varphi_{i}^{(n)}, i \in [1:L])$:
\begin{equation}
\varphi_{i}^{(n)}: \prod_{j \in [1:i]}  \mathcal{M}_{j}^{(n)} \times \mathcal{Y}_{i}^{n } \mapsto \mathcal{\hat{X}}^{n},\qquad i \in [1:L],
\end{equation}
where $\varphi_{i}^{(n)}$ produces  the source reconstruction $\hat{{X}}_{i}^{n}(M_{j}, j \in [1:i], {Y}_{i}^{n})$ by using codewords $(M_{j}, j \in [1:i])$ and side information ${Y}_{i}^{n}$. In particular, under covariance distortion contraints, there is no loss of optimality in assuming that $\phi_{i}^{(n)}$ performs MMSE estimation, i.e., $\hat{{X}}_{i}^{n}(M_{j}, j \in [1:i], {Y}_{i}^{n})=\mathbb{E}[X^n| M_{j}, j \in [1:i], {Y}_{i}^{n}]$, $i \in [1:L]$.
\end{itemize}

%\begin{definition}
%	A rate tuple $(R_{i}, i \in [1:L])$ is said to be admissible subject to distortion constraints $({D}_{i}, i \in [1:L])$, if there exists a sequence of encoding functions $(n, \phi_{i}^{(n)}, i \in [1:L])$, and decoding functions $(n, \varphi_{i}^{(n)}, i \in [1:L])$ such that
%	\begin{align}
%	\limsup_{n \rightarrow \infty}\frac{1}{n} \log \left|         \mathcal{M}_{i}^{(n)}     \right| \leq R_{i}, \qquad i \in [1:L],\\
%	\limsup_{n \rightarrow \infty} \frac{1}{n} \sum_{t=1}^{n}\mathbb{E} \left[ d_{i}(X, \hat{X}_{i}) \right] \leq D_{i}, \qquad i \in [1:L]
%	\end{align}
%	The rate-distortion region $\mathcal{R}( {D}_{i}, i \in [1:L])$ is defined as the closure of the set of all such admissible rate tuples.
%\end{definition}

\begin{definition}\it
A rate tuple $(R_{i}, i \in [1:L])$ is said to be achievable subject to covariance distortion constraints $(\boldsymbol{D}_{i} \succ \boldsymbol{0}, i \in [1:L])$ if there exists a sequence of encoding functions $(\phi_{i}^{(n)}, i \in [1:L])$ and decoding functions $(\varphi_{i}^{(n)}, i \in [1:L])$ such that
\begin{align}
\limsup_{n \rightarrow \infty}\frac{1}{n} \log \left|         \mathcal{M}_{i}^{(n)}     \right| &\leq R_{i}, \qquad i \in [1:L],\\
\limsup_{n \rightarrow \infty} \frac{1}{n} \sum_{t=1}^{n} \mathbb{E} \left[       \left( {X}(t) - {\hat{X}}_{i} (t) \right) \left( {X}(t) - {\hat{X}}_{i} (t) \right)^{T}                \right]
&\preceq \boldsymbol{D}_{i},  \qquad i \in [1:L].
\end{align}
The rate region $\mathcal{R}^*( \boldsymbol{D}_{i}, i \in [1:L])$ is defined as the set of all such achievable rate tuples.
\end{definition}

The following theorem states a computable characterization of $\mathcal{R}^*( \boldsymbol{D}_{i}, i \in [1:L])$, which is the main result of this paper.

\begin{theorem} \label{main_thm}\it
 $\mathcal{R}^*( \boldsymbol{D}_{i}, i \in [1:L])=\mathcal{R}( \boldsymbol{D}_{i}, i \in [1:L])$, where $\mathcal{R}( \boldsymbol{D}_{i}, i \in [1:L])$ is the convex hull of the set of $(R_{i}, i \in [1:L])$ such that
\begin{align}
R_{1} & \geq \frac{1}{2} \log \frac{| \boldsymbol{K}_{0}^{-1} + \boldsymbol{K}_{1}^{-1} + \boldsymbol{B}_{1} |}{|\boldsymbol{K}_{0}^{-1}+\boldsymbol{K}_{1}^{-1} |}, \\
\sum_{j=1}^{i} R_{j} & \geq \frac{1}{2} \log \frac{| \boldsymbol{K}_{0}^{-1} + \boldsymbol{K}_{1}^{-1} + \boldsymbol{B}_{1} |}{|\boldsymbol{K}_{0}^{-1}+\boldsymbol{K}_{1}^{-1} |}+ \sum_{j=2}^{i}\frac{1}{2} \log \frac{| \boldsymbol{K}_{0}^{-1} + \boldsymbol{K}_{j}^{-1} + \sum_{k=1}^{j}\boldsymbol{B}_{k}  |}{|\boldsymbol{K}_{0}^{-1}+\boldsymbol{K}_{j}^{-1}+\sum_{k=1}^{j-1}\boldsymbol{B}_{k} |}, \qquad  i\in[2:L],
\end{align}
for some $(\boldsymbol{B}_{i}, i \in [1:L])$ satisfying
\begin{align}
\boldsymbol{B}_{i} &\succeq \boldsymbol{0}, \qquad i \in[1:L], \\
\sum_{j=1}^{i} \boldsymbol{B}_{j} & \succeq \boldsymbol{D}_{i}^{-1} - \boldsymbol{K}_{0}^{-1} - \boldsymbol{K}_{i}^{-1}, \qquad i \in[1:L].
\end{align}
\end{theorem}

The proof of Theorem \ref{main_thm} can be found in Section \ref{MT}, and it relies critically on the extremal inequality established in Section~\ref{EI}.

%In Section \ref{EI}, we will derive an extremal inequality as an intermediate result. In Section \ref{MT}, it will be found that this extremal inequality plays a key role in the proof of Theorem \ref{main_thm}.

\section{An Extremal Inequality}\label{EI}

\begin{theorem}\label{ei_thm}\it
Given $\mu_{1} \geq \mu_{2} \geq \cdots \geq \mu_{L} \geq 0$, let $(\boldsymbol{B}_{i}^{*}, i \in [1:L])$ be any positive semi-definite matrices such that
\begin{equation}
\sum_{j=1}^{i} \boldsymbol{B}_{j}^{*}  \succeq \boldsymbol{D}_{i}^{-1} - \boldsymbol{K}_{0}^{-1} - \boldsymbol{K}_{i}^{-1}, \qquad i \in[1:L],
\end{equation}
and
\begin{align}
\frac{\mu_{i}}{2} \left(    \boldsymbol{K}_{0}^{-1} + \boldsymbol{K}_{i}^{-1} + \sum_{j=1}^{i} \boldsymbol{B}_{j}^{*}          \right)^{-1}- \frac{\mu_{i+1}}{2} \left(    \boldsymbol{K}_{0}^{-1} + \boldsymbol{K}_{i+1}^{-1} + \sum_{j=1}^{i} \boldsymbol{B}_{j}^{*}          \right)^{-1} &= \boldsymbol{\Psi}_{i}-\boldsymbol{\Psi}_{i+1}+\boldsymbol{\Lambda}_{i}, \qquad i \in [1:L-1], \label{id_KKT1}\\
\frac{\mu_{L}}{2} \left(    \boldsymbol{K}_{0}^{-1} + \boldsymbol{K}_{L}^{-1} + \sum_{j=1}^{L} \boldsymbol{B}_{j}^{*}          \right)^{-1} &=\boldsymbol{\Psi}_{L}+\boldsymbol{\Lambda}_{L}, \label{id_KKT2}\\
\boldsymbol{B}_{i}^{*} \boldsymbol{\Psi}_{i} &= \boldsymbol{0}, \qquad i \in [1:L], \label{CS_IQ}\\
\left( \boldsymbol{K}_{0}^{-1} + \boldsymbol{K}_{i}^{-1} + \sum_{j=1}^{i} \boldsymbol{B}_{j}^{*} - \boldsymbol{D}_{i}^{-1} \right) \boldsymbol{\Lambda}_{i} &= \boldsymbol{0}, \qquad i \in [1:L], \label{CS_eq}
\end{align}
for some positive semi-definite matrices $(\boldsymbol{\Psi}_{i}, i \in [1:L])$ and $(\boldsymbol{\Lambda}_{i}, i \in [1:L])$.
%\begin{align}
%\boldsymbol{B}_{i}^{*} \boldsymbol{\Psi}_{i} &= \boldsymbol{0}, \qquad i \in [1:L], \label{CS_IQ}\\
%\left( \boldsymbol{K}_{0}^{-1} + \boldsymbol{K}_{i}^{-1} + \sum_{j=1}^{i} \boldsymbol{B}_{j}^{*} - \boldsymbol{D}_{i}^{-1} \right) \boldsymbol{\Lambda}_{i} &= \boldsymbol{0}, \qquad i \in [1:L]. \label{CS_eq}
%\end{align}
For any random objects $(W_{i}, i \in [1:L])$ satisfying the Markov chain contraint
\begin{equation}
(W_{i}, i \in [1:L]) \rightarrow X\rightarrow Y_{L} \rightarrow Y_{L-1} \rightarrow \ldots \rightarrow Y_{1}
\end{equation}
and the covariance distortion constraints
\begin{align}
\cov(X|Y_{i}, W_{j}, j \in [1:i]) \preceq \boldsymbol{D}_{i}, \qquad i \in [1:L], \label{dist}
\end{align}
the following extremal inequality holds:
\begin{align}\label{ex_inq}
&\sum_{i=1}^{L-1}    \left(    \mu_{i}h(Y_{i}| W_{j}, j \in [1:i] ) - \mu_{i+1}h(Y_{i+1}| W_{j}, j \in [1:i]) - (\mu_{i} - \mu_{i+1})h(X | W_{j}, j \in [1:i])                     \right)      \nonumber \\
& + \mu_{L} h(Y_{L}| W_{j}, j \in [1:L]) - \mu_{L} h(X | W_{j}, j \in [1:L]) \nonumber \\
& \geq \sum_{i=1}^{L-1}  \left(  -\frac{\mu_{i+1}}{2} \log \left| (2 \pi e)^{-1}\left( \boldsymbol{K}_{0}^{-1} + \boldsymbol{K}_{i+1}^{-1} + \sum_{j=1}^{i} \boldsymbol{B}_{j}^{*}   \right)  \right|  + \frac{\mu_{i}}{2} \log \left| (2 \pi e)^{-1}\left(    \boldsymbol{K}_{0}^{-1} + \boldsymbol{K}_{i}^{-1} + \sum_{j=1}^{i} \boldsymbol{B}_{j}^{*}          \right)  \right| \right)\nonumber \\
& \quad +  \frac{\mu_{L}}{2} \log \left| (2 \pi e)^{-1} \left(    \boldsymbol{K}_{0}^{-1} + \boldsymbol{K}_{L}^{-1} + \sum_{j=1}^{L} \boldsymbol{B}_{j}^{*}          \right)  \right|.
\end{align}
%$\mu_{i+1}h(X|Y_{i+1}, W_{j}, j \in [1:i])-\mu_{i}h(X|Y_{i}, W_{j}, j \in [1:i])$
%\begin{align}\label{ex_inq}
%&\sum_{i=1}^{L-1}    \left((\mu_{i}- \mu_{i+1})h(X|Y_{i+1}, W_{j}, j \in [1:i])+ \mu_{i}h(Y_{i+1}|Y_{i}, W_{j}, j \in [1:i]) \right)       + \mu_{L} h(X|Y_{L}, W_{j}, j \in [1:L]) \nonumber \\
%& \leq \sum_{i=1}^{L-1}  \left(  \frac{\mu_{i+1}}{2} \log \left| (2 \pi e)^{-1}\left( \boldsymbol{K}_{0}^{-1} + \boldsymbol{K}_{i+1}^{-1} + \sum_{j=1}^{i} \boldsymbol{B}_{j}^{*}   \right)  \right| - \frac{\mu_{i}}{2} \log \left| (2 \pi e)^{-1}\left(    \boldsymbol{K}_{0}^{-1} + \boldsymbol{K}_{i}^{-1} + \sum_{j=1}^{i} \boldsymbol{B}_{j}^{*}          \right)  \right| \right.\nonumber \\
%& \qquad \qquad \left. + \frac{\mu_{i}}{2} \log \left| (2 \pi e)\left( \boldsymbol{K}_{i} - \boldsymbol{K}_{i+1} \right) \right| +   \frac{\mu_{i}}{2} \log \left| (2 \pi e) \boldsymbol{K}_{i} \right|  - \frac{\mu_{i}}{2} \log \left| (2 \pi e) \boldsymbol{K}_{i+1}   \right|   \right)\nonumber \\
%& \quad -  \frac{\mu_{L}}{2} \log \left| (2 \pi e)^{-1} \left(    \boldsymbol{K}_{0}^{-1} + \boldsymbol{K}_{L}^{-1} + \sum_{j=1}^{L} \boldsymbol{B}_{j}^{*}          \right)  \right|.
%\end{align}
\end{theorem}
%\begin{remark}
%When $L=2$ under the Markov chain assumption:
%\begin{equation}
%(W_{1},W_{2}) \rightarrow X \rightarrow (Y_{1},Y_{2}),
%\end{equation}
%by setting $\boldsymbol{\Psi}_{2}=\boldsymbol{\Lambda}_{2} = \boldsymbol{0}$, the extremal inequality \eqref{ex_inq} can be decomposed into the following two extremal inequality:
%\begin{align}
%- \mu_{2} h(X|Y_{2},W_{1},W_{2}) &\leq \frac{\mu_{2}}{2} \log \left| (2 \pi e) \left(    \boldsymbol{K}_{0}^{-1} + \boldsymbol{K}_{2}^{-1} + \boldsymbol{B}_{1}^{*}  + \boldsymbol{B}_{2}^{*}          \right)  \right|, \label{ex_one} \\
% \mu_{2} h(X|Y_{2},W_{1}) -  \mu_{1} h(X|Y_{1},W_{1})  &\leq - \frac{\mu_{2}}{2} \log \left| (2 \pi e)\left( \boldsymbol{K}_{0}^{-1} + \boldsymbol{K}_{2}^{-1}+ \boldsymbol{B}_{1}^{*}   \right)  \right|  + \frac{\mu_{1}}{2} \log \left| (2 \pi e)\left(    \boldsymbol{K}_{0}^{-1} + \boldsymbol{K}_{1}^{-1} +  \boldsymbol{B}_{1}^{*} \right)  \right|. \label{ex_two}
%\end{align}
%\eqref{ex_one} can be obtained by worst additive noise lemma \cite[Lemma II.2]{DC01}, and \eqref{ex_two} can be regarded as a variant of the extremal inequality motivated by the vector Gaussian secret key generation in \cite[Theorem 3]{WO11}, the proof of which is to exploit the degraded orders between vector Gaussian sources, by using the seminal enhancement argument \cite{WSS06}.
%\end{remark}
\begin{remark}
For the special case $L=2$, $\boldsymbol{\Lambda}_{1} =\boldsymbol{0}$, and $\mu_{1}=\mu_{2}=1$, the extremal inequality \eqref{ex_inq} can be regarded as a variant of  \cite[Theorem 5]{EU13}, the original proof of which relies on the enhancement argument developed in \cite{WSS06}. However, when $L> 2$, the enhancement argument appears to be inadequate for resolving the difficulty caused by the introduction of
$(\boldsymbol{\Psi}_{i}, i \in [1:L])$ and $(\boldsymbol{\Lambda}_{i}, i \in [1:L])$. We shall overcome this difficulty via a judicious application of the monotone path argument and the doubling trick.

%However, for the extremal inequality \eqref{ex_inq} with $L\geq 2$ under the Markov chain assumption:
%\begin{equation}
%(W_{i}, i \in [1:L]) \rightarrow X\rightarrow Y_{L} \rightarrow Y_{L-1} \rightarrow \ldots \rightarrow Y_{1},
%\end{equation}
%the enhancement argument seems insufficient to resolve the difficulties arising when introducing $(\boldsymbol{\Psi}_{i}, i \in [1:L])$ and $(\boldsymbol{\Lambda}_{i}, i \in [1:L])$. In this paper, we develop a new Gaussian perturbation approach. By making use of the degraded side information assumption, we further obtain the new extremal inequality \eqref{ex_inq}.
\end{remark}

%\subsection{Proof of Theorem \ref{ei_thm}}

For notational simplicity, we define \begin{equation}\label{eq:simp}
\boldsymbol{\Delta}_{i}^{-1}\triangleq  \boldsymbol{K}_{0}^{-1} + \sum_{j=1}^{i} \boldsymbol{B}_{j}^{*}, \qquad i \in [1:L].
\end{equation}
The proof of Theorem \ref{ei_thm} is divided into four steps.

%In the rest part of this section, we prove the extremal inequality \eqref{ex_inq} as stated in Theorem \ref{ei_thm}. At the beginning, we denote
%\begin{equation}\label{eq:simp}
%\boldsymbol{\Delta}_{i}^{-1}\triangleq  \boldsymbol{K}_{0}^{-1} + \sum_{j=1}^{i} \boldsymbol{B}_{j}^{*}, \qquad i \in [1:L] ,
%\end{equation}
%for the sake of compactness of notations. Next, we consider the perturbation approach to \eqref{ex_inq}. The proof is rather long so we divide it into the following four steps.

\subsection{Constructing the Monotone Path}
We first construct $3L$ zero-mean Gaussian random vectors
\begin{equation*}
X_{1}^{G}, \ldots, X_{L}^{G},\; Y_{1}^{G}, \ldots, Y_{L}^{G}, \; \tilde{Y}_{2}^{G}, \ldots, \tilde{Y}_{L+1}^{G},
\end{equation*}
which are independent of $(X_{i}, Y_{i}, W_{i}, i \in [1:L])$. Specifically, they are defined as follows.

\subsubsection{}
 Let $X_{L}^{G}$, $W^G_i$, $i\in[2:L]$, be mutually independent Gaussian random vectors with covariance matrices $\boldsymbol{\Delta}_{L}$, $\boldsymbol{\Delta}_{i-1} - \boldsymbol{\Delta}_{i}$, $i\in[2:L]$, respectively. We define
\begin{equation}
X^{G}_{i} = X^{G}_{i+1} + W^{G}_{i+1}, \qquad i \in [1:L-1].
\end{equation}
It is easy to see that
\begin{equation}
X^{G}_{i} \sim \mathcal{N}\left(\boldsymbol{0}, \boldsymbol{\Delta}_{i}\right), \qquad i \in [1:L].
\end{equation}

\subsubsection{}
 Let $N_{i}^{G}-N_{i+1}^G$, $i\in[1:L],$ be mutually independent Gaussian random vectors with covariance matrices $\boldsymbol{K}_{i}-\boldsymbol{K}_{i+1}$, $i\in[1:L]$, respectively. We assume that $(N^G_i, i\in[1:L+1])$ is independent of $(X^G_i, i\in[1:L])$. Define
\begin{align}
&Y_{i}^{G}  = X_{i}^{G} + N_{i}^{G}, \qquad i \in [1:L],\label{eq:TYG}\\
&\tilde{Y}_{i}^{G}  = X_{i-1}^{G} + N_{i}^{G}, \qquad i \in [2:L+1]. \label{eq:YG}
\end{align}
It is clear that
\begin{align}
&Y_{i}^{G}  \sim \mathcal{N}\left(\boldsymbol{0}, \boldsymbol{\Delta}_{i} + \boldsymbol{K}_{i}\right), \qquad i \in [1:L],\\
&\tilde{Y}_{i}^{G}  \sim \mathcal{N}\left(\boldsymbol{0}, \boldsymbol{\Delta}_{i-1} + \boldsymbol{K}_{i}\right), \qquad i \in [2:L+1].
\end{align}

 Using the covariance preserved transform (see, e.g., \cite{DCT91}), we define
  \begin{align}
&{X}_{i,\gamma}  =  \sqrt{1-\gamma}X + \sqrt{\gamma}{X}_{i}^{G}, \qquad i \in[1:L],\label{eqn:per1}\\
%{X}^{*}_{i,\gamma} & = \sqrt{\gamma}{X} -  \sqrt{1-\gamma} {X}_{i}^{G}, \qquad i \in [1:L],\label{eqn:per2}\\
&{Y}_{i,\gamma} = \sqrt{1-\gamma}Y_{i} + \sqrt{\gamma}{Y}_{i}^{G}, \qquad i \in[1:L],\label{eqn:per3}\\
&\tilde{Y}_{i, \gamma} = \sqrt{1-\gamma}Y_{i}+ \sqrt{\gamma}{\tilde{Y}}_{i}^{G}, \qquad i \in[2:L+1]\footnotemark[2],\label{eqn:per4}\\
&Y^{*}_{i,\gamma} = \sqrt{\gamma}{Y}_{i} -  \sqrt{1-\gamma} {Y}_{i}^{G}, \qquad i \in[1:L], \label{eqn:per5}
\end{align}
for any $\gamma \in (0,1)$.
\footnotetext[2]{Here $Y_{L+1} = X$ and $\tilde{Y}_{L+1,\gamma} = X_{L,\gamma}$.}
Consider the following function:
\begin{align}
g(\gamma)=& \sum_{i=1}^{L-1}    \Big(    \mu_{i}h(Y_{i,\gamma}|Y^{*}_{i,\gamma}, W_{j}, j \in [1:i] ) - \mu_{i+1}h(\tilde{Y}_{i+1,\gamma}|Y^{*}_{i,\gamma}, W_{j}, j \in [1:i]) \nonumber\\
& \qquad - (\mu_{i} - \mu_{i+1})h(X_{i,\gamma} |Y^{*}_{i,\gamma}, W_{j}, j \in [1:i])                     \Big)      \nonumber \\
& \quad + \mu_{L} h(Y_{L,\gamma}|Y^{*}_{L,\gamma}, W_{j}, j \in [1:L]) - \mu_{L} h(X_{L,\gamma} |Y^{*}_{L,\gamma}, W_{j}, j \in [1:L]). \label{eq:fung}
%& \left( \mu_{i+1}h\left(X_{i, \gamma}\Big|\tilde{Y}_{i+1, \gamma},{X}^{*}_{i, \gamma}, W_{j}, j \in [1:i]\right)-\mu_{i}h\left(X_{i, \gamma}\Big|Y_{i, \gamma},{X}^{*}_{i, \gamma}, W_{j}, j \in [1:i]\right) \right)    \nonumber \\
\end{align}
Notice that $g(0)$ coincides with the left-hand side of \eqref{ex_inq} while $g(1)$ coincides with the right-hand side of \eqref{ex_inq}. Therefore, it suffices to show that $g(\gamma)$ decreases monotonically along the path parameterized by $\gamma$, i.e.,
\begin{equation}\label{ineq:dev1}
\frac{d}{d\gamma}g(\gamma) \leq 0, \qquad \gamma\in(0,1).
\end{equation}

%\begin{align}
%g(\gamma)= &\sum_{i=1}^{L-1}    \left( \mu_{i+1}h\left(X_{i, \gamma}\Big|\tilde{Y}_{i+1, \gamma},{X}^{*}_{i, \gamma}, W_{j}, j \in [1:i]\right)-\mu_{i}h\left(X_{i, \gamma}\Big|Y_{i, \gamma},\tilde{X}_{i, \gamma}, W_{j}, j \in [1:i]\right) \right)    \nonumber \\
%& \quad  - \mu_{L}h\left(X_{L, \gamma}\Big|Y_{L, \gamma},\tilde{X}_{L, \gamma}, W_{j}, j \in [1:L]\right) \label{eq:fung}
%\end{align}

\begin{remark}
The construction of $({Y}^{*}_{i, \gamma}, i \in [1:L])$ is inspired by the doubling trick introduced in \cite{GN14}. A similar construction can be found in \cite{WC19}.
\end{remark}

\subsection{Derivative of $g(\gamma)$}
In this step, we utilize a vector generalization of I-MMSE relationship from \cite{GSV05}. First rewrite \eqref{eq:fung} as
\begin{align}
g(\gamma)= &\sum_{i=1}^{L-1} \Big(    \mu_{i}h(Y_{i,\gamma}, Y^{*}_{i,\gamma}| W_{j}, j \in [1:i] ) - \mu_{i+1}h(\tilde{Y}_{i+1,\gamma}, Y^{*}_{i,\gamma}| W_{j}, j \in [1:i]) \nonumber \\
& \qquad - (\mu_{i} - \mu_{i+1})h(X_{i,\gamma}, Y^{*}_{i,\gamma}| W_{j}, j \in [1:i])                     \Big)      \nonumber \\
& + \mu_{L} h(Y_{L,\gamma}, Y^{*}_{L,\gamma}| W_{j}, j \in [1:L]) - \mu_{L} h(X_{L,\gamma}, Y^{*}_{L,\gamma}|  W_{j}, j \in [1:L]).
\end{align}
In view of \eqref{eqn:per3} and \eqref{eqn:per5}, it can be verified that
\begin{align}
&h\left(Y_{i, \gamma}, {Y}^{*}_{i, \gamma} \Big| W_{j}, j \in [1:i]\right)\nonumber \\
&= h\left(    \sqrt{1-\gamma}Y_{i} + \sqrt{\gamma}{Y}_{i}^{G},    \sqrt{\gamma}{Y}_{i} -  \sqrt{1-\gamma} {Y}_{i}^{G}  \Big| W_{j}, j \in [1:i]  \right)\\
&= h\left(    Y_{i},   {Y}_{i}^{G}   \Big| W_{j}, j \in [1:i]  \right), \qquad i \in [1:L]. \label{eqn:tev1}
\end{align}
Since $Y_{i}$ and ${Y}_{i}^{G}$ do not depend on $\gamma$, it follows that
\begin{equation}
\frac{d}{d\gamma} h\left(    Y_{i},   {Y}_{i}^{G}   \Big| W_{j}, j \in [1:i]  \right) = 0,  \qquad i \in [1:L].\label{eqn:tev2}
\end{equation}
Moreover, as shown in Appendices \ref{app_tev1} and \ref{app_tev2},
\begin{align}
&\frac{d}{d\gamma}h\left({X}_{i, \gamma}, {Y}^{*}_{i, \gamma}\Big| W_{j}, j \in [1:i]\right) \nonumber \\
&=\frac{1}{2(1-\gamma)} \tr \left\{   (\boldsymbol{\Delta}_{i}^{-1}+ \boldsymbol{K}_{i}^{-1} )^{-1} \left(  J\left(  {X}_{i, \gamma}\Big |  {Y}^{*}_{i, \gamma},  W_{j}, j \in [1:i]   \right) \ - \boldsymbol{\Delta}_{i}^{-1} \right)  \right\}, \qquad i \in [1:L] . \label{eqn:tev3} \\
&\frac{d}{d\gamma}h\left(\tilde{Y}_{i+1, \gamma} , {Y}^{*}_{i, \gamma}\Big| W_{j}, j \in [1:i]\right) \nonumber \\
&=\frac{1}{2(1-\gamma)} \tr \Big\{   \left(  \left( \boldsymbol{\Delta}_{i}^{-1}+\boldsymbol{K}_{i}^{-1}           \right)^{-1} - \left( \boldsymbol{\Delta}_{i}^{-1}+\boldsymbol{K}_{i+1}^{-1}           \right)^{-1}   \right)\Big( \left( \boldsymbol{\Delta}_{i}^{-1}+\boldsymbol{K}_{i+1}^{-1}           \right)\boldsymbol{K}_{i+1} \nonumber\\
& \qquad J\left(  \tilde{Y}_{i+1, \gamma} \Big |  {Y}^{*}_{i, \gamma},  W_{j}, j \in [1:i]   \right)\boldsymbol{K}_{i+1} \left( \boldsymbol{\Delta}_{i}^{-1}+\boldsymbol{K}_{i+1}^{-1}           \right)\ - \boldsymbol{\Delta}_{i}^{-1} \left(\boldsymbol{\Delta}_{i}+\boldsymbol{K}_{i+1}\right) \boldsymbol{\Delta}_{i}^{-1}\Big)  \Big\}      , \qquad i \in [1:L-1].\label{eqn:tev4}
\end{align}
Combining \eqref{eqn:tev1}, \eqref{eqn:tev2}, \eqref{eqn:tev3}, and \eqref{eqn:tev4} gives
\begin{align}
& -2(1-\gamma) \frac{d}{d\gamma}g(\gamma)\nonumber \\
&= \sum_{i=1}^{L-1} \tr \left\{   \left(  \mu_{i+1}\left( \boldsymbol{\Delta}_{i}^{-1}+\boldsymbol{K}_{i}^{-1}           \right)^{-1} - \mu_{i+1}\left( \boldsymbol{\Delta}_{i}^{-1}+\boldsymbol{K}_{i+1}^{-1}           \right)^{-1}   \right)\left( \left( \boldsymbol{\Delta}_{i}^{-1}+\boldsymbol{K}_{i+1}^{-1}           \right)\boldsymbol{K}_{i+1} \right.\right.\nonumber\\
& \qquad \qquad \quad \left.\left.J\left(  \tilde{Y}_{i+1, \gamma} \Big |  {Y}^{*}_{i, \gamma},  W_{j}, j \in [1:i]   \right)\boldsymbol{K}_{i+1} \left( \boldsymbol{\Delta}_{i}^{-1}+\boldsymbol{K}_{i+1}^{-1}           \right)\ - \boldsymbol{\Delta}_{i}^{-1} \left(\boldsymbol{\Delta}_{i}+\boldsymbol{K}_{i+1}\right) \boldsymbol{\Delta}_{i}^{-1}\right)  \right\} \nonumber \\
& \quad + \sum_{i=1}^{L-1} \tr \left\{ (\mu_{i}-\mu_{i+1})  (\boldsymbol{\Delta}_{i}^{-1}+ \boldsymbol{K}_{i}^{-1} )^{-1} \left(  J\left(  {X}_{i, \gamma}\Big |  {Y}^{*}_{i, \gamma},  W_{j}, j \in [1:i]   \right) \ -\boldsymbol{\Delta}_{i}^{-1} \right)  \right\}\nonumber \\
& \quad + \tr \left\{   \mu_{L}(\boldsymbol{\Delta}_{L}^{-1}+ \boldsymbol{K}_{L}^{-1} ) \left(  J\left(  {X}_{L, \gamma}\Big |  {Y}^{*}_{L, \gamma},  W_{j}, j \in [1:L]   \right) \ - \boldsymbol{\Delta}_{L}^{-1} \right)  \right\},\qquad\gamma\in(0,1). \label{ieq:lbb}
\end{align}
Hence, for the purpose of proving \eqref{ineq:dev1}, it suffices to show that \eqref{ieq:lbb} is greater than or equal to $0$.

\subsection{Lower Bound of \eqref{ieq:lbb}}

%In order to represent our extremal inequality in the form of incorporating the  Karush-Kuhn-Tucker (KKT) conditions in \eqref{id_KKT1} and \eqref{id_KKT2}, we give a lower bound of \eqref{ieq:lbb} in the next subsection.

In this step, we establish a lower bound of \eqref{ieq:lbb} with the  Karush-Kuhn-Tucker (KKT) conditions in \eqref{id_KKT1} and \eqref{id_KKT2}  properly incorporated.
First notice that the covariance matrix of random vector
$$
\left(\begin{array}{r}
 \sqrt{1-\gamma}N_{i+1}+ \sqrt{\gamma}N_{i+1}^{G}\\
\sqrt{\gamma}N_{i}- \sqrt{1-\gamma}N_{i}^{G}
\end{array}\right)
$$
is given by
\begin{equation}
 \begin{pmatrix}
\boldsymbol{K}_{i+1} & \boldsymbol{0}\\
\boldsymbol{0} & \boldsymbol{K}_{i}
\end{pmatrix}.
\end{equation}
So $\sqrt{1-\gamma}N_{i+1}+ \sqrt{\gamma}N_{i+1}^{G}$ is independent of $\sqrt{\gamma}N_{i}- \sqrt{1-\gamma}N_{i}^{G}$, which, together with  \eqref{eqn:per5}, implies that $\sqrt{1-\gamma}N_{i+1}+ \sqrt{\gamma}N_{i+1}^{G}$ is independent of $Y_{i,\gamma}^{*}$ as well.
For $i \in [1:L-1]$, we have
\begin{equation}
\tilde{Y}_{i+1,\gamma} = X_{i,\gamma}+\sqrt{1-\gamma}N_{i+1}+ \sqrt{\gamma}N_{i+1}^{G}.
\end{equation}
 In view of the fact that $\sqrt{1-\gamma}N_{i+1}+ \sqrt{\gamma}N_{i+1}^{G}$ is independent of $X_{i,\gamma}$,  the Fisher information inequality (see Lemma \ref{fi_inq} in Appendix \ref{app_lea2}) can be invoked to show
\begin{align}
 &\left( \boldsymbol{\Delta}_{i}^{-1}+\boldsymbol{K}_{i+1}^{-1}           \right)\boldsymbol{K}_{i+1} J\left(  \tilde{Y}_{i+1, \gamma} \Big |  {Y}^{*}_{i, \gamma},  W_{j}, j \in [1:i]   \right)\boldsymbol{K}_{i+1} \left( \boldsymbol{\Delta}_{i}^{-1}+\boldsymbol{K}_{i+1}^{-1}           \right)\ - \boldsymbol{\Delta}_{i}^{-1} \left(\boldsymbol{\Delta}_{i}+\boldsymbol{K}_{i+1}\right) \boldsymbol{\Delta}_{i}^{-1} \nonumber \\
 &=\left(\boldsymbol{I}+ \boldsymbol{\Delta}_{i}^{-1}\boldsymbol{K}_{i+1}\right)J\left( X_{i,\gamma}+\sqrt{1-\gamma}N_{i+1}+ \sqrt{\gamma}N_{i+1}^{G} \Big |  {Y}^{*}_{i, \gamma},  W_{j}, j \in [1:i]   \right)\left( \boldsymbol{K}_{i+1}\boldsymbol{\Delta}_{i}^{-1}+\boldsymbol{I}\right) \nonumber\\
 &\quad-\boldsymbol{\Delta}_{i}^{-1}\boldsymbol{K}_{i+1}\boldsymbol{\Delta}_{i}^{-1}-\boldsymbol{\Delta}_{i}^{-1} \\
 & \preceq J\left( X_{i,\gamma} \Big |  {Y}^{*}_{i, \gamma},  W_{j}, j \in [1:i]   \right) + \boldsymbol{\Delta}_{i}^{-1}\boldsymbol{K}_{i+1}J\left( \sqrt{1-\gamma}N_{i+1}+ \sqrt{\gamma}N_{i+1}^{G} \right)\boldsymbol{K}_{i+1}\boldsymbol{\Delta}_{i}^{-1}\nonumber \\
 & \quad -\boldsymbol{\Delta}_{i}^{-1}\boldsymbol{K}_{i+1}\boldsymbol{\Delta}_{i}^{-1}-\boldsymbol{\Delta}_{i}^{-1}\\
 %&=J\left( X_{i,\gamma} \Big |  {Y}^{*}_{i, \gamma},  W_{j}, j \in [1:i]   \right)-\frac{1}{\gamma}\boldsymbol{\Delta}_{i}^{-1} -\frac{1-\gamma}{\gamma}\boldsymbol{\Delta}_{i}^{-1}\boldsymbol{K}_{i+1}\boldsymbol{\Delta}_{i}^{-1} \\
 &= J\left( X_{i,\gamma} \Big |  {Y}^{*}_{i, \gamma},  W_{j}, j \in [1:i]   \right)-\boldsymbol{\Delta}_{i}^{-1}.
\end{align}
Since $\boldsymbol{K}_{i}\succ \boldsymbol{K}_{i+1}$, it follows that
\begin{equation}
\left( \boldsymbol{\Delta}_{i}^{-1}+\boldsymbol{K}_{i}^{-1}           \right)^{-1} - \left( \boldsymbol{\Delta}_{i}^{-1}+\boldsymbol{K}_{i+1}^{-1}           \right)^{-1}\succ \boldsymbol{0}.
\end{equation}
Therefore,
\begin{align}
& -2(1-\gamma) \frac{d}{d\gamma}g(\gamma)\nonumber \\
%&\leq \sum_{i=1}^{L-1} \tr \left\{   \left(  \mu_{i+1}\left( \boldsymbol{\Delta}_{i}^{-1}+\boldsymbol{K}_{i}^{-1}           \right)^{-1} - \mu_{i+1}\left( \boldsymbol{\Delta}_{i}^{-1}+\boldsymbol{K}_{i+1}^{-1}           \right)^{-1}   \right)\left(        J\left(  {X}_{i, \gamma}\Big |  {Y}^{*}_{i, \gamma},  W_{j}, j \in [1:i]   \right) \ - \frac{1}{\gamma}\boldsymbol{\Delta}_{i}^{-1}       \right)  \right\} \nonumber \\
%& \quad + \sum_{i=1}^{L-1} \tr \left\{ (\mu_{i}-\mu_{i+1})  (\boldsymbol{\Delta}_{i}^{-1}+ \boldsymbol{K}_{i}^{-1} )^{-1} \left(  J\left(  {X}_{i, \gamma}\Big |  {Y}^{*}_{i, \gamma},  W_{j}, j \in [1:i]   \right) \ - \frac{1}{\gamma}\boldsymbol{\Delta}_{i}^{-1} \right)  \right\}\nonumber \\
%& \quad - \tr \left\{   \mu_{L}(\boldsymbol{\Delta}_{L}^{-1}+ \boldsymbol{K}_{L}^{-1} ) \left(  J\left(  {X}_{L, \gamma}\Big |  {Y}^{*}_{L, \gamma},  W_{j}, j \in [1:L]   \right) \ - \frac{1}{\gamma}\boldsymbol{\Delta}_{L}^{-1} \right)  \right\} \\
&\geq \sum_{i=1}^{L-1} \tr \Big\{   \left(  \mu_{i}\left( \boldsymbol{\Delta}_{i}^{-1}+\boldsymbol{K}_{i}^{-1}           \right)^{-1} - \mu_{i+1}\left( \boldsymbol{\Delta}_{i}^{-1}+\boldsymbol{K}_{i+1}^{-1}           \right)^{-1}   \right)\Big( \left( \boldsymbol{\Delta}_{i}^{-1}+\boldsymbol{K}_{i+1}^{-1}           \right)\boldsymbol{K}_{i+1} \nonumber\\
& \qquad \qquad \quad J\left(  \tilde{Y}_{i+1, \gamma} \Big |  {Y}^{*}_{i, \gamma},  W_{j}, j \in [1:i]   \right)\boldsymbol{K}_{i+1} \left( \boldsymbol{\Delta}_{i}^{-1}+\boldsymbol{K}_{i+1}^{-1}           \right)\ - \boldsymbol{\Delta}_{i}^{-1} \left(\boldsymbol{\Delta}_{i}+\boldsymbol{K}_{i+1}\right) \boldsymbol{\Delta}_{i}^{-1}\Big)  \Big\} \nonumber\\
&\quad - \tr \left\{   \mu_{L}(\boldsymbol{\Delta}_{L}^{-1}+ \boldsymbol{K}_{L}^{-1} ) \left(  J\left(  {X}_{L, \gamma}\Big |  {Y}^{*}_{L, \gamma},  W_{j}, j \in [1:L]   \right) \ - \boldsymbol{\Delta}_{L}^{-1} \right)  \right\} \\
&=\sum_{i=1}^{L-1}\tr \Big\{   \left(  \boldsymbol{\Psi}_{i}-\boldsymbol{\Psi}_{i+1}+\boldsymbol{\Lambda}_{i}  \right)\Big( \left( \boldsymbol{\Delta}_{i}^{-1}+\boldsymbol{K}_{i+1}^{-1}           \right)\boldsymbol{K}_{i+1} \nonumber\\
& \qquad \qquad \quad  J\left(  \tilde{Y}_{i+1, \gamma} \Big |  {Y}^{*}_{i, \gamma},  W_{j}, j \in [1:i]   \right)\boldsymbol{K}_{i+1} \left( \boldsymbol{\Delta}_{i}^{-1}+\boldsymbol{K}_{i+1}^{-1}           \right)\ - \boldsymbol{\Delta}_{i}^{-1} \left(\boldsymbol{\Delta}_{i}+\boldsymbol{K}_{i+1}\right) \boldsymbol{\Delta}_{i}^{-1}\Big)  \Big\} \nonumber \\
& \quad + \tr \left\{  \left( \boldsymbol{\Psi}_{L}+\boldsymbol{\Lambda}_{L} \right)\left(  J\left(  \tilde{Y}_{L+1, \gamma}\Big |  {Y}^{*}_{L, \gamma},  W_{j}, j \in [1:L]   \right) \ - \boldsymbol{\Delta}_{L}^{-1} \right)  \right\}, \label{ex_inq1}
\end{align}
where \eqref{ex_inq1} is due to the KKT properties in \eqref{id_KKT1} and \eqref{id_KKT2}. Via suitable rearrangement, this lower bound can be written in the following equivalent form:
\begin{subequations}
\begin{align}
& -2(1-\gamma) \frac{d}{d\gamma}g(\gamma)\nonumber \\
&\geq \tr \left\{    \boldsymbol{\Psi}_{1}\left( \left( \boldsymbol{\Delta}_{1}^{-1}+\boldsymbol{K}_{2}^{-1}           \right)\boldsymbol{K}_{2}J\left(  \tilde{Y}_{2, \gamma} \Big |  {Y}^{*}_{1, \gamma},  W_{1}   \right)\boldsymbol{K}_{2} \left( \boldsymbol{\Delta}_{1}^{-1}+\boldsymbol{K}_{2}^{-1}           \right)\ - \boldsymbol{\Delta}_{1}^{-1} \left(\boldsymbol{\Delta}_{1}+\boldsymbol{K}_{2}\right) \boldsymbol{\Delta}_{1}^{-1}\right)  \right\} \label{ex_main1} \\
&\quad+\sum_{i=2}^{L} \tr \Big\{     \boldsymbol{\Psi}_{i}\Big(\left( \boldsymbol{\Delta}_{i}^{-1}+\boldsymbol{K}_{i+1}^{-1}           \right)\boldsymbol{K}_{i+1} J\left(  \tilde{Y}_{i+1, \gamma} \Big |  {Y}^{*}_{i, \gamma},  W_{j}, j \in [1:i]   \right)\boldsymbol{K}_{i+1} \left( \boldsymbol{\Delta}_{i}^{-1}+\boldsymbol{K}_{i+1}^{-1}           \right)  \nonumber\\
& \qquad \qquad \quad \left.\left.-\left( \boldsymbol{\Delta}_{i-1}^{-1}+\boldsymbol{K}_{i}^{-1}           \right)\boldsymbol{K}_{i}J\left(  \tilde{Y}_{i, \gamma} \Big |  {Y}^{*}_{i-1, \gamma},  W_{j}, j \in [1:i-1]   \right)\boldsymbol{K}_{i} \left( \boldsymbol{\Delta}_{i-1}^{-1}+\boldsymbol{K}_{i}^{-1}           \right) \right.\right.\nonumber\\
 & \qquad \qquad \quad - \boldsymbol{\Delta}_{i}^{-1} \left(\boldsymbol{\Delta}_{i}+\boldsymbol{K}_{i+1}\right) \boldsymbol{\Delta}_{i}^{-1}+ \boldsymbol{\Delta}_{i-1}^{-1} \left(\boldsymbol{\Delta}_{i-1}+\boldsymbol{K}_{i}\right) \boldsymbol{\Delta}_{i-1}^{-1}\Big)  \Big\} \label{ex_main2} \\
&\quad+\sum_{i=1}^{L}\tr \Big\{   \boldsymbol{\Lambda}_{i} \Big( \left( \boldsymbol{\Delta}_{i}^{-1}+\boldsymbol{K}_{i+1}^{-1}           \right)\boldsymbol{K}_{i+1}J\left(  \tilde{Y}_{i+1, \gamma} \Big |  {Y}^{*}_{i, \gamma},  W_{j}, j \in [1:i]   \right)\boldsymbol{K}_{i+1} \left( \boldsymbol{\Delta}_{i}^{-1}+\boldsymbol{K}_{i+1}^{-1}           \right) \nonumber\\
& \qquad \qquad \quad \ - \boldsymbol{\Delta}_{i}^{-1} \left(\boldsymbol{\Delta}_{i}+\boldsymbol{K}_{i+1}\right) \boldsymbol{\Delta}_{i}^{-1}\Big)   \Big\}. \label{ex_main3}
\end{align}
\end{subequations}
Now it suffices to show that \eqref{ex_main1}-\eqref{ex_main3} are all lower bounded by 0.  %Henceforth, the rest part of this section is devoted to derive the lower bound of \eqref{ex_main1}-\eqref{ex_main3}, separately.

\subsection{Lower Bound of \eqref{ex_main1}}
From \eqref{eq:tt2.1} in Appendix \ref{app_tev2},
\begin{align}
& \left( \boldsymbol{\Delta}_{1}^{-1}+\boldsymbol{K}_{2}^{-1}           \right)\boldsymbol{K}_{2}J\left(  \tilde{Y}_{2, \gamma} \Big |  {Y}^{*}_{1, \gamma},  W_{1}   \right)\boldsymbol{K}_{2} \left( \boldsymbol{\Delta}_{1}^{-1}+\boldsymbol{K}_{2}^{-1}           \right)\ - \boldsymbol{\Delta}_{1}^{-1} \left(\boldsymbol{\Delta}_{1}+\boldsymbol{K}_{2}\right) \boldsymbol{\Delta}_{1}^{-1} \nonumber \\
& = \frac{1-\gamma}{\gamma}\boldsymbol{\Delta}_{1}^{-1}\left(\boldsymbol{\Delta}_{1}+\boldsymbol{K}_{2}\right)\left(    \left(\boldsymbol{\Delta}_{1}+\boldsymbol{K}_{2}\right)^{-1} +\left(\boldsymbol{K}_{1}-\boldsymbol{K}_{2}\right)^{-1}   \right)
\left( \left(    \left(\boldsymbol{\Delta}_{1}+\boldsymbol{K}_{2}\right)^{-1} +\left(\boldsymbol{K}_{1}-\boldsymbol{K}_{2}\right)^{-1}   \right)^{-1}    \right. \nonumber \\
& \quad \left.  - \frac{1}{\gamma}  \cov \left(   Y_{2} \Big | \tilde{Y}_{2, \gamma}, {Y}^{*}_{1, \gamma}, W_{1}     \right)   \right)
\left(    \left(\boldsymbol{\Delta}_{1}+\boldsymbol{K}_{2}\right)^{-1} +\left(\boldsymbol{K}_{1}-\boldsymbol{K}_{2}\right)^{-1}   \right)\left(\boldsymbol{\Delta}_{1}+\boldsymbol{K}_{2}\right)\boldsymbol{\Delta}_{1}^{-1} \label{inq:Q1}.
\end{align}
Combining the data processing inequality for MMSE (see Lemma \ref{DP_MMSE} in Appendix \ref{app_lea2}) and \eqref{eq_152}  gives
\begin{align}
\cov \left(   Y_{2} \Big | \tilde{Y}_{2, \gamma}, {Y}^{*}_{1, \gamma}, W_{1}     \right) &\preceq \cov \left(   Y_{2} \Big | \tilde{Y}_{2, \gamma}, {Y}^{*}_{1, \gamma}   \right) \nonumber \\
&=\left(  \left(\boldsymbol{K}_{0}+ \boldsymbol{K}_{2}\right)^{-1}+  \frac{1-\gamma}{\gamma} \left(\boldsymbol{\Delta}_{1} + \boldsymbol{K}_{2}\right)^{-1} + \frac{1}{\gamma}\left(  \boldsymbol{K}_{1} - \boldsymbol{K}_{2}\right)^{-1}         \right)^{-1}.\label{eq:lowerQ1}
\end{align}
Substituting \eqref{eq:lowerQ1} into \eqref{inq:Q1} yields the following lower bound:
\begin{align}
& \left( \boldsymbol{\Delta}_{1}^{-1}+\boldsymbol{K}_{2}^{-1}           \right)\boldsymbol{K}_{2}J\left(  \tilde{Y}_{2, \gamma} \Big |  {Y}^{*}_{1, \gamma},  W_{1}   \right)\boldsymbol{K}_{2} \left( \boldsymbol{\Delta}_{1}^{-1}+\boldsymbol{K}_{2}^{-1}           \right)\ - \boldsymbol{\Delta}_{1}^{-1} \left(\boldsymbol{\Delta}_{1}+\boldsymbol{K}_{2}\right) \boldsymbol{\Delta}_{1}^{-1} \nonumber \\
&\succeq \frac{(1-\gamma)^{2}}{\gamma^{2}}\boldsymbol{\Delta}_{1}^{-1}\left(\boldsymbol{\Delta}_{1}+\boldsymbol{K}_{2}\right)\left(   \left(\boldsymbol{\Delta}_{1}+\boldsymbol{K}_{2}\right)^{-1} +\left(\boldsymbol{K}_{1}-\boldsymbol{K}_{2}\right)^{-1}   \right)
\Big(  \frac{\gamma}{1-\gamma} \left(    \left(\boldsymbol{\Delta}_{1}+\boldsymbol{K}_{2}\right)^{-1} +\left(\boldsymbol{K}_{1}-\boldsymbol{K}_{2}\right)^{-1}   \right)^{-1}    \nonumber \\
& \quad   - \frac{1}{1-\gamma}  \Big(  \left(\boldsymbol{K}_{0}+ \boldsymbol{K}_{2}\right)^{-1}+  \frac{1-\gamma}{\gamma} \left(\boldsymbol{\Delta}_{1} + \boldsymbol{K}_{2}\right)^{-1} + \frac{1}{\gamma}\left(  \boldsymbol{K}_{1} - \boldsymbol{K}_{2}\right)^{-1}         \Big)^{-1}   \Big) \nonumber \\
& \quad \left(    \left(\boldsymbol{\Delta}_{1}+\boldsymbol{K}_{2}\right)^{-1} +\left(\boldsymbol{K}_{1}-\boldsymbol{K}_{2}\right)^{-1}   \right)\left(\boldsymbol{\Delta}_{1}+\boldsymbol{K}_{2}\right)\boldsymbol{\Delta}_{1}^{-1} \\
& = \frac{1-\gamma}{\gamma}\boldsymbol{\Delta}_{1}^{-1}\left(\boldsymbol{\Delta}_{1}+\boldsymbol{K}_{2}\right)\left(   \left(\boldsymbol{\Delta}_{1}+\boldsymbol{K}_{2}\right)^{-1} +\left(\boldsymbol{K}_{1}-\boldsymbol{K}_{2}\right)^{-1}   \right)\nonumber \\
& \quad \left(  \left(\boldsymbol{K}_{0}+ \boldsymbol{K}_{2}\right)^{-1}+  \frac{1-\gamma}{\gamma} \left(\boldsymbol{\Delta}_{1} + \boldsymbol{K}_{2}\right)^{-1} + \frac{1}{\gamma}\left(  \boldsymbol{K}_{1} - \boldsymbol{K}_{2}\right)^{-1}         \right)^{-1}\left(     \left(\boldsymbol{K}_{0}+ \boldsymbol{K}_{2}\right)^{-1} -    \left(\boldsymbol{\Delta}_{1} + \boldsymbol{K}_{2}\right)^{-1}   \right)\left(\boldsymbol{\Delta}_{1}+\boldsymbol{K}_{2}\right)\boldsymbol{\Delta}_{1}^{-1} \\
&=\frac{1-\gamma}{\gamma}\boldsymbol{\Delta}_{1}^{-1}\left(\boldsymbol{\Delta}_{1}+\boldsymbol{K}_{2}\right)\left(   \left(\boldsymbol{\Delta}_{1}+\boldsymbol{K}_{2}\right)^{-1} +\left(\boldsymbol{K}_{1}-\boldsymbol{K}_{2}\right)^{-1}   \right)\nonumber \\
& \quad \left(  \left(\boldsymbol{K}_{0}+ \boldsymbol{K}_{2}\right)^{-1}+  \frac{1-\gamma}{\gamma} \left(\boldsymbol{\Delta}_{1} + \boldsymbol{K}_{2}\right)^{-1} + \frac{1}{\gamma}\left(  \boldsymbol{K}_{1} - \boldsymbol{K}_{2}\right)^{-1}         \right)^{-1}\left(\boldsymbol{K}_{0}+ \boldsymbol{K}_{2}\right)^{-1}\left(     \boldsymbol{\Delta}_{1} - \boldsymbol{K}_{0}  \right)\boldsymbol{\Delta}_{1}^{-1} \\
&=\frac{1-\gamma}{\gamma}\boldsymbol{\Delta}_{1}^{-1}\left(\boldsymbol{\Delta}_{1}+\boldsymbol{K}_{2}\right)\left(   \left(\boldsymbol{\Delta}_{1}+\boldsymbol{K}_{2}\right)^{-1} +\left(\boldsymbol{K}_{1}-\boldsymbol{K}_{2}\right)^{-1}   \right)\nonumber \\
& \quad \left(  \left(\boldsymbol{K}_{0}+ \boldsymbol{K}_{2}\right)^{-1}+  \frac{1-\gamma}{\gamma} \left(\boldsymbol{\Delta}_{1} + \boldsymbol{K}_{2}\right)^{-1} + \frac{1}{\gamma}\left(  \boldsymbol{K}_{1} - \boldsymbol{K}_{2}\right)^{-1}         \right)^{-1}\left(\boldsymbol{K}_{0}+ \boldsymbol{K}_{2}\right)^{-1}\boldsymbol{K}_{0}^{-1}\left(      \boldsymbol{K}_{0}^{-1} -  \boldsymbol{\Delta}_{1}^{-1}  \right).
\end{align}
From the complementary slackness condition in \eqref{CS_IQ}, i.e.,
\begin{align}
\boldsymbol{B}^{*}_{1} \boldsymbol{\Psi}_{1} = \left( \boldsymbol{K}^{-1}_{0} - \boldsymbol{\Delta}^{-1}_{1}  \right) \boldsymbol{\Psi}_{1} &= \boldsymbol{0},
\end{align}
we have
\begin{align}
&\tr \left\{    \boldsymbol{\Psi}_{1}\left( \left( \boldsymbol{\Delta}_{1}^{-1}+\boldsymbol{K}_{2}^{-1}           \right)\boldsymbol{K}_{2}J\left(  \tilde{Y}_{2, \gamma} \Big |  {Y}^{*}_{1, \gamma},  W_{1}   \right)\boldsymbol{K}_{2} \left( \boldsymbol{\Delta}_{1}^{-1}+\boldsymbol{K}_{2}^{-1}           \right)\ - \boldsymbol{\Delta}_{1}^{-1} \left(\boldsymbol{\Delta}_{1}+\boldsymbol{K}_{2}\right) \boldsymbol{\Delta}_{1}^{-1}\right)  \right\} \\
& \geq \tr \Big\{    \frac{1-\gamma}{\gamma}\boldsymbol{\Delta}_{1}^{-1}\left(\boldsymbol{\Delta}_{1}+\boldsymbol{K}_{2}\right)\left(   \left(\boldsymbol{\Delta}_{1}+\boldsymbol{K}_{2}\right)^{-1} +\left(\boldsymbol{K}_{1}-\boldsymbol{K}_{2}\right)^{-1}   \right)\nonumber \\
& \qquad \quad \Big(  \left(\boldsymbol{K}_{0}+ \boldsymbol{K}_{2}\right)^{-1}+  \frac{1-\gamma}{\gamma} \left(\boldsymbol{\Delta}_{1} + \boldsymbol{K}_{2}\right)^{-1} + \frac{1}{\gamma}\left(  \boldsymbol{K}_{1} - \boldsymbol{K}_{2}\right)^{-1}         \Big)^{-1}\left(\boldsymbol{K}_{0}+ \boldsymbol{K}_{2}\right)^{-1}\boldsymbol{K}_{0}^{-1}\left(      \boldsymbol{K}_{0}^{-1} -  \boldsymbol{\Delta}_{1}^{-1}  \right)\boldsymbol{\Psi}_{1}                     \Big\} \nonumber \\
&= 0.
\end{align}
This proves that \eqref{ex_main1} is lower bounded by 0.

\subsection{Lower Bound of \eqref{ex_main2}}
To the end of showing that \eqref{ex_main2} is lower bounded by 0, we introduce
\begin{equation}
{N}{'}_{i+1} \triangleq \sqrt{1-\gamma}\left(N_{i}-N_{i+1}\right)+\sqrt{\gamma}\left(N_{i}^{G}-N_{i+1}^{G}\right), \qquad i \in [1:L]. \label{eq:nota}
\end{equation}
Note that ${N}'_{i+1}$ is a Gaussian random vector with covariance matrix $\boldsymbol{K}_{i} - \boldsymbol{K}_{i+1}$ and is independent of $(\tilde{Y}_{i+1,\gamma}, {Y}^{*}_{i, \gamma})$. Moreover,
\begin{equation}
\tilde{Y}_{i,\gamma} = \tilde{Y}_{i+1,\gamma}+{N}'_{i+1},\qquad i \in [2:L].
\end{equation}
In view of the fact that ${N}'_{i+1}$ is independent of ${Y}^{*}_{i, \gamma}$, we can  invoke the Fisher information inequality (see  Lemma \ref{fi_inq} in Appendix \ref{app_lea2}) to show
\begin{align}
&\left( \boldsymbol{\Delta}_{i-1}^{-1}+\boldsymbol{K}_{i}^{-1}           \right)\boldsymbol{K}_{i}J\left(  \tilde{Y}_{i, \gamma} \Big |  {Y}^{*}_{i-1, \gamma},  W_{j}, j \in [1:i-1]   \right)\boldsymbol{K}_{i} \left( \boldsymbol{K}_{i}^{-1}+\boldsymbol{\Delta}_{i-1}^{-1}           \right) \nonumber\\
&=\left( \boldsymbol{\Delta}_{i-1}^{-1}\boldsymbol{K}_{i} +\boldsymbol{I}           \right) J\left(  \tilde{Y}_{i+1, \gamma}+N'_{i+1} \Big |  {Y}^{*}_{i-1, \gamma},  W_{j}, j \in [1:i-1]   \right) \left(\boldsymbol{I}+\boldsymbol{K}_{i} \boldsymbol{\Delta}_{i-1}^{-1}           \right) \\
& \preceq \left( \boldsymbol{\Delta}_{i-1}^{-1}\boldsymbol{K}_{i+1} +\boldsymbol{I}           \right)J\left(  \tilde{Y}_{i+1, \gamma} \Big |  {Y}^{*}_{i-1, \gamma},  W_{j}, j \in [1:i-1]   \right)\left(\boldsymbol{I}+\boldsymbol{K}_{i+1} \boldsymbol{\Delta}_{i-1}^{-1}           \right)+ \boldsymbol{\Delta}_{i-1}^{-1} \left(    \boldsymbol{K}_{i}- \boldsymbol{K}_{i+1}    \right)\boldsymbol{\Delta}_{i-1}^{-1}\\
& \preceq  \left( \boldsymbol{\Delta}_{i-1}^{-1}\boldsymbol{K}_{i+1} +\boldsymbol{I}           \right)J\left(  \tilde{Y}_{i+1, \gamma} \Big |  {Y}^{*}_{i, \gamma},  W_{j}, j \in [1:i]   \right)\left(\boldsymbol{I}+\boldsymbol{K}_{i+1} \boldsymbol{\Delta}_{i-1}^{-1}           \right)+ \boldsymbol{\Delta}_{i-1}^{-1} \left(    \boldsymbol{K}_{i}- \boldsymbol{K}_{i+1}    \right)\boldsymbol{\Delta}_{i-1}^{-1} \label{eq:be1}\\
&  = \left( \boldsymbol{\Delta}_{i-1}^{-1} +\boldsymbol{K}_{i+1}^{-1}          \right)\boldsymbol{K}_{i+1} J\left(  \tilde{Y}_{i+1, \gamma} \Big |  {Y}^{*}_{i, \gamma},  W_{j}, j \in [1:i]   \right) \boldsymbol{K}_{i+1}\left(\boldsymbol{K}_{i+1}^{-1}+ \boldsymbol{\Delta}_{i-1}^{-1}           \right)+\boldsymbol{\Delta}_{i-1}^{-1} \left(    \boldsymbol{K}_{i}- \boldsymbol{K}_{i+1}    \right)\boldsymbol{\Delta}_{i-1}^{-1},\label{eq:be2}
\end{align}
where \eqref{eq:be1} follows by the Markov chain contraint $ \left(    {Y}^{*}_{i-1, \gamma},  W_{j}, j \in [1:i-1]   \right) \rightarrow \left(  {Y}^{*}_{i, \gamma},  W_{j}, j \in [1:i]      \right)  \rightarrow \tilde{Y}_{i+1, \gamma}$ and the data processing inequality for Fisher information (see  Lemma \ref{DP_FI} in Appendix \ref{app_lea2}). %, and \eqref{eq:be2} is because of $\gamma \in (0,1)$.
Meanwhile, due to the complementary slackness condition in \eqref{CS_IQ}, i.e.,
\begin{align}
\boldsymbol{B}^{*}_{i} \boldsymbol{\Psi}_{i} = \left( \boldsymbol{\Delta}^{-1}_{i} - \boldsymbol{\Delta}^{-1}_{i-1}  \right) \boldsymbol{\Psi}_{i} &= \boldsymbol{0}, \qquad i \in [2:L],
\end{align}
we have
\begin{align}
&\tr \Big\{     \boldsymbol{\Psi}_{i}\Big(\left( \boldsymbol{\Delta}_{i}^{-1}+\boldsymbol{K}_{i+1}^{-1}           \right)\boldsymbol{K}_{i+1} J\left(  \tilde{Y}_{i+1, \gamma} \Big |  {Y}^{*}_{i, \gamma},  W_{j}, j \in [1:i]   \right)\boldsymbol{K}_{i+1} \left( \boldsymbol{\Delta}_{i}^{-1}+\boldsymbol{K}_{i+1}^{-1}           \right)  \nonumber\\
& \qquad \qquad \quad \left.\left.-\left( \boldsymbol{\Delta}_{i-1}^{-1}+\boldsymbol{K}_{i}^{-1}           \right)\boldsymbol{K}_{i}J\left(  \tilde{Y}_{i, \gamma} \Big |  {Y}^{*}_{i-1, \gamma},  W_{j}, j \in [1:i-1]   \right)\boldsymbol{K}_{i} \left( \boldsymbol{\Delta}_{i-1}^{-1}+\boldsymbol{K}_{i}^{-1}           \right) \right.\right.\nonumber\\
 & \qquad \qquad \quad - \boldsymbol{\Delta}_{i}^{-1} \left(\boldsymbol{\Delta}_{i}+\boldsymbol{K}_{i+1}\right) \boldsymbol{\Delta}_{i}^{-1}+ \boldsymbol{\Delta}_{i-1}^{-1} \left(\boldsymbol{\Delta}_{i-1}+\boldsymbol{K}_{i}\right) \boldsymbol{\Delta}_{i-1}^{-1}\Big)  \Big\} \nonumber \\
 &= \tr \Big\{     \boldsymbol{\Psi}_{i}\Big(\left( \boldsymbol{\Delta}_{i-1}^{-1}+\boldsymbol{K}_{i+1}^{-1}           \right)\boldsymbol{K}_{i+1} J\left(  \tilde{Y}_{i+1, \gamma} \Big |  {Y}^{*}_{i, \gamma},  W_{j}, j \in [1:i]   \right)\boldsymbol{K}_{i+1} \left( \boldsymbol{\Delta}_{i-1}^{-1}+\boldsymbol{K}_{i+1}^{-1}           \right)  \nonumber\\
& \qquad \qquad \quad \left.\left.-\left( \boldsymbol{\Delta}_{i-1}^{-1}+\boldsymbol{K}_{i}^{-1}           \right)\boldsymbol{K}_{i}J\left(  \tilde{Y}_{i, \gamma} \Big |  {Y}^{*}_{i-1, \gamma},  W_{j}, j \in [1:i-1]   \right)\boldsymbol{K}_{i} \left( \boldsymbol{\Delta}_{i-1}^{-1}+\boldsymbol{K}_{i}^{-1}           \right) \right.\right.\nonumber\\
 & \qquad \qquad \quad+ \boldsymbol{\Delta}_{i-1}^{-1} \left( \boldsymbol{K}_{i}-\boldsymbol{K}_{i+1}       \right)\boldsymbol{\Delta}_{i-1}^{-1} \Big)  \Big\} \nonumber\\
 &\geq 0, \qquad i \in [2:L].
\end{align}
This proves that \eqref{ex_main2} is lower bounded by 0.

\subsection{Lower Bound of \eqref{ex_main3}}
%From the perturbation variable constructions of $(\tilde{Y}_{i+1, \gamma}, {Y}^{*}_{i, \gamma})$ in \eqref{eqn:per4} and \eqref{eqn:per5}, we write the following inequalities on MMSE,
%\begin{align}
%&\cov \left(   Y_{i+1} \Big | \tilde{Y}_{i+1, \gamma}, {Y}^{*}_{i, \gamma}, W_{j}, j \in [1:i]     \right) \nonumber \\
%&=\cov \left(  Y_{i+1} \Big |  \sqrt{1-\gamma}Y_{i+1}+\sqrt{\gamma}\tilde{Y}_{i+1}^{G}, \sqrt{\gamma}Y_{i} - \sqrt{1-\gamma}Y_{i}^{G}, W_{j}, j \in [1:i]       \right)\\
%&=\cov \left(  Y_{i+1} \Big |  \sqrt{1-\gamma}Y_{i+1}+\sqrt{\gamma}\tilde{Y}_{i+1}^{G}, \sqrt{\gamma}Y_{i} - \sqrt{1-\gamma}Y_{i}^{G}, W_{j}, j \in [1:i]       \right)
%\end{align}
%
%
%we shall represent $N_{i}-N_{i+1}$ by its linear estimation when giving $\left( \sqrt{\gamma}\tilde{Y}_{i+1}^{G}, \sqrt{\gamma}\left(N_{i}-N_{i+1}\right) - \sqrt{1-\gamma}Y_{i}^{G} \right)$
%
%
%
%
%
%
%
%\newpage

%Moreover, by the complementary identity in Lemma \ref{lea1}, we give the Fisher information representation of the MMSE terms in \eqref{eqn:main1}.

To the end of showing that \eqref{ex_main3} is lower bounded by 0, we introduce
\begin{equation}
{N}{''}_{i+1} \triangleq \sqrt{\gamma}\left(N_{i}-N_{i+1}\right)-\sqrt{1-\gamma}\left(N_{i}^{G}-N_{i+1}^{G}\right), \qquad i \in [1:L].
\end{equation}
Note that ${N}''_{i+1}$ is a Gaussian random vector with covariance matrix $\boldsymbol{K}_{i} - \boldsymbol{K}_{i+1}$ and is independent of $({Y}_{i+1}, \tilde{Y}_{i+1}^{G})$.
It can be verified that
\begin{align}
&\cov \left(   Y_{i+1} \Big | \tilde{Y}_{i+1, \gamma}, {Y}^{*}_{i, \gamma}, W_{j}, j \in [1:i]     \right) \nonumber \\
&= \cov \left(    Y_{i+1} \Big |\sqrt{1-\gamma}\tilde{Y}_{i+1, \gamma}+\sqrt{\gamma}{Y}^{*}_{i, \gamma},  \tilde{Y}_{i+1, \gamma}, W_{j}, j \in [1:i]                                  \right) \\
&=\cov \left(   Y_{i+1} \Big | (1-\gamma)Y_{i+1}+\sqrt{\gamma(1-\gamma)}\tilde{Y}_{i+1}^{G}+\gamma Y_{i} - \sqrt{\gamma(1-\gamma)}Y_{i}^{G} ,  \tilde{Y}_{i+1, \gamma}, W_{j}, j \in [1:i]     \right)\\
&=\cov \left(   Y_{i+1} \Big |   Y_{i+1}+\sqrt{\gamma}{N}''_{i+1} , Y_{i+1}+\sqrt{\frac{\gamma}{1-\gamma}}\tilde{Y}_{i+1}^{G},             W_{j}, j \in [1:i]                \right)\\
%&=\gamma\left(\boldsymbol{K}_{i} - \boldsymbol{K}_{i+1}\right)-\gamma\left(\boldsymbol{K}_{i} - \boldsymbol{K}_{i+1}\right)J\left(Y_{i+1}+\sqrt{\gamma}{N}'_{i+1} | W_{j}, j \in [1:i] \right)\gamma\left(\boldsymbol{K}_{i} - \boldsymbol{K}_{i+1}\right). \label{eq:zhu3}
&\preceq  \cov \left(   Y_{i+1} \Big |  \left(       \frac{1-\gamma}{\gamma}  \left( \boldsymbol{\Delta}_{i} + \boldsymbol{K}_{i+1}\right)^{-1}+ \frac{1}{\gamma} \left( \boldsymbol{K}_{i} - \boldsymbol{K}_{i+1}\right)^{-1}     \right)          Y_{i+1}      +   \sqrt{\frac{1-\gamma}{\gamma}}  \left( \boldsymbol{\Delta}_{i} + \boldsymbol{K}_{i+1}\right)^{-1} \tilde{Y}_{i+1}^{G}     \right.\nonumber \\
& \qquad \qquad \left.  +  \sqrt{\frac{1}{\gamma}} \left( \boldsymbol{K}_{i} - \boldsymbol{K}_{i+1}\right)^{-1}  {N}''_{i+1},  W_{j}, j \in [1:i]                  \right).  \label{eq:zhu3}
\end{align}
where \eqref{eq:zhu3} is due to the data processing inequality for MMSE (see Lemma \ref{DP_MMSE} in Appendix \ref{app_lea2}).

Let
\begin{align}
\boldsymbol{P}_{i+1}\triangleq & \left(       \frac{1-\gamma}{\gamma}  \left( \boldsymbol{\Delta}_{i} + \boldsymbol{K}_{i+1}\right)^{-1}+ \frac{1}{\gamma} \left( \boldsymbol{K}_{i} - \boldsymbol{K}_{i+1}\right)^{-1}\right)^{-1}, \\
S^{G}_{i+1} \triangleq & \boldsymbol{P}_{i+1}\left(\sqrt{\frac{1-\gamma}{\gamma}}  \left( \boldsymbol{\Delta}_{i} + \boldsymbol{K}_{i+1}\right)^{-1} \tilde{Y}_{i+1}^{G}+  \sqrt{\frac{1}{\gamma}} \left( \boldsymbol{K}_{i} - \boldsymbol{K}_{i+1}\right)^{-1}  {N}''_{i+1}\right).
\end{align}
It follows by the theory of linear MMSE estimtion that
\begin{align}
N_{i}-N_{i+1} = S_{i+1}^{G} + T_{i+1}^{G},
\end{align}
where $T_{i+1}^{G}$ is a Gaussian random vector with covariance matrix
$
 \boldsymbol{K}_{i} - \boldsymbol{K}_{i+1}-\boldsymbol{P}_{i+1},
$
and is independent of $S_{i+1}^{G}$.

Due to the Markov chain
\begin{equation}
\left( W_{j}, j \in [1:i] \right)\rightarrow Y_{i+1} \rightarrow   Y_{i+1}+S_{i+1}^{G} \rightarrow Y_{i+1}+S_{i+1}^{G}+T_{i+1}^{G},
\end{equation}
we can invoke Lemma \ref{MMSE_inq} in Appendix \ref{app_lea2} to show that
\begin{align}
&\cov \left(   Y_{i+1} \Big | \tilde{Y}_{i+1, \gamma}, {Y}^{*}_{i, \gamma}, W_{j}, j \in [1:i]     \right)^{-1} \\
&\succeq \cov \left(   Y_{i+1} \Big | Y_{i+1}+S_{i+1}^{G}, W_{j}, j \in [1:i]     \right)^{-1}  \\
& \succeq \cov \left(   Y_{i+1} \Big | Y_{i+1}+S_{i+1}^{G}+T_{i+1}^{G}, W_{j}, j \in [1:i]     \right)^{-1} - \left(\boldsymbol{K}_{i} - \boldsymbol{K}_{i+1}\right)^{-1} + \boldsymbol{P}_{i+1}^{-1} \\
& = \cov \left(   Y_{i+1} \Big | Y_{i}, W_{j}, j \in [1:i]     \right)^{-1} + \frac{1-\gamma}{\gamma} \left(         \left( \boldsymbol{\Delta}_{i} + \boldsymbol{K}_{i+1}\right)^{-1}+  \left( \boldsymbol{K}_{i} - \boldsymbol{K}_{i+1}\right)^{-1}\right).
\end{align}

\subsubsection{}
Note that the following Markov chain condition holds:
\begin{equation}\label{con_markov}
(W_{j}, j \in [1:i]) \rightarrow X \rightarrow Y_{i+1} \rightarrow Y_{i}.
\end{equation}
Since $X$, $Y_{i}$, and $Y_{i+1})$ are jointly Gaussian, it follows that
\begin{align}
&\mathbb{E} \left[ Y_{i+1}| X, Y_{i} \right] \\
%&  = \mathbb{E} \left[ X+N_{i+1}| X, X+N_{i} \right] \\
%&  = \mathbb{E} \left[ X+N_{i+1}| X, N_{i} \right] \\
%&  = \mathbb{E} \left[ X| X, N_{i} \right]+\mathbb{E} \left[N_{i+1}| X, N_{i} \right]\\
%&  = X + \mathbb{E} \left[N_{i+1}| N_{i} \right] \label{equ1}\\
%&  = X+   \boldsymbol{K}_{i+1}  \boldsymbol{K}_{i}^{-1}   N_{i} \\
&  =  \left( \boldsymbol{K}_{i} - \boldsymbol{K}_{i+1} \right)\boldsymbol{K}_{i}^{-1}  X + \boldsymbol{K}_{i+1}  \boldsymbol{K}_{i}^{-1} Y_{i}.\label{equ2}
\end{align}
%where \eqref{equ1} holds because $(N_i, N_{i+1})$ and $X$ are independent, and \eqref{equ2} follows from $Y_i=X+N_i$.
Furthermore, we have
\begin{align}\label{eq:mark}
Y_{i+1}= & \left( \boldsymbol{K}_{i} - \boldsymbol{K}_{i+1} \right)\boldsymbol{K}_{i}^{-1} \left(  X + \tilde{N}_{i+1}       \right)+ \boldsymbol{K}_{i+1}  \boldsymbol{K}_{i}^{-1} Y_{i},
\end{align}
where $\tilde{N}_{i+1}$ is a zero-mean Gaussian random vector with  covariance matrix
\begin{align}\label{equ5}
\tilde{\boldsymbol{K}}_{i+1}= \left( \boldsymbol{K}_{i+1}^{-1} - \boldsymbol{K}_{i}^{-1} \right)^{-1}\succ \boldsymbol{0},
\end{align}
and is independent of $(X, Y_{i})$.
Therefore,
\begin{align}
&\cov \left(   Y_{i+1} \Big | Y_{i}, W_{j}, j \in [1:i]     \right) \nonumber \\
&= \cov \left(   \left( \boldsymbol{K}_{i} - \boldsymbol{K}_{i+1} \right)\boldsymbol{K}_{i}^{-1} \left(  X + \tilde{N}_{i+1}       \right) \Big | Y_{i}, W_{j}, j \in [1:i]     \right) \\
%&= \left( \boldsymbol{K}_{i} - \boldsymbol{K}_{i+1} \right)\boldsymbol{K}_{i}^{-1} \cov \left(     X   \Big | Y_{i}, W_{j}, j \in [1:i]   \right)\boldsymbol{K}_{i}^{-1}\left( \boldsymbol{K}_{i} - \boldsymbol{K}_{i+1} \right) + \boldsymbol{K}_{i+1} \left( \boldsymbol{K}^{-1}_{i+1} - \boldsymbol{K}^{-1}_{i}        \right) \boldsymbol{K}_{i+1}  \\
&\preceq \left( \boldsymbol{K}_{i} - \boldsymbol{K}_{i+1} \right)\boldsymbol{K}_{i}^{-1} \left(\boldsymbol{D}_{i} + \left( \boldsymbol{K}_{i+1}^{-1} - \boldsymbol{K}_{i}^{-1} \right)^{-1} \right)\boldsymbol{K}_{i}^{-1}\left( \boldsymbol{K}_{i} - \boldsymbol{K}_{i+1} \right), \label{inq79}
\end{align}
where \eqref{inq79} is because of covariance distortion constraint $\cov(X|Y_{i}, W_{j}, j \in [1:i]) \preceq \boldsymbol{D}_{i}$ in \eqref{dist}.

\subsubsection{}
It can be verified that
\begin{align}
&\left(         \left( \boldsymbol{\Delta}_{i} + \boldsymbol{K}_{i+1}\right)^{-1}+  \left( \boldsymbol{K}_{i} - \boldsymbol{K}_{i+1}\right)^{-1}\right)^{-1} \nonumber \\
&=\left(          \boldsymbol{K}_{i+1}^{-1} -    \left( \boldsymbol{K}_{i+1} - \boldsymbol{K}_{i}\right)^{-1}  -\boldsymbol{K}_{i+1}^{-1}    +  \left( \boldsymbol{\Delta}_{i} + \boldsymbol{K}_{i+1}\right)^{-1}                 \right)^{-1} \\
&=\boldsymbol{K}_{i+1} \left(   \left( \boldsymbol{K}^{-1}_{i+1} - \boldsymbol{K}^{-1}_{i}\right)^{-1}   -      \left( \boldsymbol{\Delta}^{-1}_{i} + \boldsymbol{K}^{-1}_{i+1}\right)^{-1}               \right)^{-1} \boldsymbol{K}_{i+1}\\
&=\boldsymbol{K}_{i+1}\left( \boldsymbol{K}^{-1}_{i+1} - \boldsymbol{K}^{-1}_{i}\right) \left(    \left( \boldsymbol{\Delta}^{-1}_{i} + \boldsymbol{K}^{-1}_{i}\right)^{-1}  +     \left( \boldsymbol{K}^{-1}_{i+1} - \boldsymbol{K}^{-1}_{i}\right)^{-1}             \right) \left( \boldsymbol{K}^{-1}_{i+1} - \boldsymbol{K}^{-1}_{i}\right)\boldsymbol{K}_{i+1}\\
& \preceq  \left( \boldsymbol{K}_{i} - \boldsymbol{K}_{i+1} \right)\boldsymbol{K}_{i}^{-1} \left(\boldsymbol{D}_{i} + \left( \boldsymbol{K}_{i+1}^{-1} - \boldsymbol{K}_{i}^{-1} \right)^{-1} \right)\boldsymbol{K}_{i}^{-1}\left( \boldsymbol{K}_{i} - \boldsymbol{K}_{i+1} \right)\label{eq80},
\end{align}
where \eqref{eq80} is because of  $\boldsymbol{\Delta}^{-1}_{i} + \boldsymbol{K}^{-1}_{i} \succeq \boldsymbol{D}^{-1}_{i}$.

Combing \eqref{inq79} and \eqref{eq80} yields
\begin{align}
&\cov \left(   Y_{i+1} \Big | \tilde{Y}_{i+1, \gamma}, {Y}^{*}_{i, \gamma}, W_{j}, j \in [1:i]     \right) \nonumber \\
&\preceq \gamma \left( \boldsymbol{K}_{i} - \boldsymbol{K}_{i+1} \right)\boldsymbol{K}_{i}^{-1} \left(\boldsymbol{D}_{i} + \left( \boldsymbol{K}_{i+1}^{-1} - \boldsymbol{K}_{i}^{-1} \right)^{-1} \right)\boldsymbol{K}_{i}^{-1}\left( \boldsymbol{K}_{i} - \boldsymbol{K}_{i+1} \right).
\end{align}
In view of \eqref{eq:tt2.1}, we have
\begin{align}
& \left( \boldsymbol{\Delta}_{i}^{-1}+\boldsymbol{K}_{i+1}^{-1}           \right)\boldsymbol{K}_{i+1}J\left(  \tilde{Y}_{i+1, \gamma} \Big |  {Y}^{*}_{i, \gamma},  W_{j}, j \in [1:i]   \right)\boldsymbol{K}_{i+1}\left( \boldsymbol{\Delta}_{i}^{-1}+\boldsymbol{K}_{i+1}^{-1}           \right)  - \boldsymbol{\Delta}_{i}^{-1} \left(\boldsymbol{\Delta}_{i}+\boldsymbol{K}_{i+1}\right) \boldsymbol{\Delta}_{i}^{-1}\nonumber \\
&=  \boldsymbol{\Delta}^{-1}_{i}\left( \boldsymbol{\Delta}_{i}+\boldsymbol{K}_{i+1}          \right) \left(  J\left(  \tilde{Y}_{i+1, \gamma} \Big |  {Y}^{*}_{i, \gamma},  W_{j}, j \in [1:i]   \right)- \left( \boldsymbol{\Delta}_{i}+\boldsymbol{K}_{i+1}\right)^{-1}         \right) \left( \boldsymbol{\Delta}_{i}+\boldsymbol{K}_{i+1}\right)  \boldsymbol{\Delta}^{-1}_{i} \\
&=\frac{1-\gamma}{\gamma}\boldsymbol{\Delta}^{-1}_{i}\left( \boldsymbol{\Delta}_{i}+\boldsymbol{K}_{i+1}          \right)\left(         \left( \boldsymbol{\Delta}_{i} + \boldsymbol{K}_{i+1}\right)^{-1}+  \left( \boldsymbol{K}_{i} - \boldsymbol{K}_{i+1}\right)^{-1}\right)\nonumber \\
& \qquad \left(   \left(         \left( \boldsymbol{\Delta}_{i} + \boldsymbol{K}_{i+1}\right)^{-1}+  \left( \boldsymbol{K}_{i} - \boldsymbol{K}_{i+1}\right)^{-1}\right)^{-1} - \frac{1}{\gamma}  \cov \left(   Y_{i+1} \Big | \tilde{Y}_{i+1, \gamma}, {Y}^{*}_{i, \gamma}, W_{j}, j \in [1:i]     \right)     \right)\nonumber \\
& \qquad \left(         \left( \boldsymbol{\Delta}_{i} + \boldsymbol{K}_{i+1}\right)^{-1}+  \left( \boldsymbol{K}_{i} - \boldsymbol{K}_{i+1}\right)^{-1}\right)\left( \boldsymbol{\Delta}_{i}+\boldsymbol{K}_{i+1}          \right)\boldsymbol{\Delta}^{-1}_{i}\\
& \succeq \frac{1-\gamma}{\gamma}\left( \boldsymbol{\Delta}^{-1}_{i}+\boldsymbol{K}^{-1}_{i}          \right)\left(         \left( \boldsymbol{\Delta}^{-1}_{i} + \boldsymbol{K}^{-1}_{i}\right)^{-1} -\boldsymbol{D}_{i}    \right)\left( \boldsymbol{\Delta}^{-1}_{i}+\boldsymbol{K}^{-1}_{i}          \right)\\
& = \frac{1-\gamma}{\gamma}\left( \boldsymbol{\Delta}^{-1}_{i}+\boldsymbol{K}^{-1}_{i}          \right)\boldsymbol{D}_{i}\left(\boldsymbol{D}^{-1}_{i}-\boldsymbol{\Delta}^{-1}_{i}-\boldsymbol{K}^{-1}_{i} \right).
\end{align}
From the complementary slackness condition in \eqref{CS_eq}, i.e.,
\begin{align}
\left( \boldsymbol{\Delta}^{-1}_{i}+\boldsymbol{K}^{-1}_{i} - \boldsymbol{D}_{i}^{-1} \right) \boldsymbol{\Lambda}_{i} &= \boldsymbol{0}, \qquad i \in[1:L],
\end{align}
we have
\begin{align}
&\tr \left\{   \boldsymbol{\Lambda}_{i} \left( \left( \boldsymbol{\Delta}_{i}^{-1}+\boldsymbol{K}_{i+1}^{-1}           \right)\boldsymbol{K}_{i+1}J\left(  \tilde{Y}_{i+1, \gamma} \Big |  {Y}^{*}_{i, \gamma},  W_{j}, j \in [1:i]   \right)\boldsymbol{K}_{i+1} \left( \boldsymbol{\Delta}_{i}^{-1}+\boldsymbol{K}_{i+1}^{-1}           \right) - \frac{1}{\gamma}\boldsymbol{\Delta}_{i}^{-1} \left(\boldsymbol{\Delta}_{i}+\boldsymbol{K}_{i+1}\right) \boldsymbol{\Delta}_{i}^{-1}\right)\right\} \\
& \geq \tr \left\{    \frac{1-\gamma}{\gamma} \boldsymbol{\Lambda}_{i} \left( \boldsymbol{\Delta}^{-1}_{i}+\boldsymbol{K}^{-1}_{i}          \right)\boldsymbol{D}_{i}\left(\boldsymbol{D}^{-1}_{i}-\boldsymbol{\Delta}^{-1}_{i}-\boldsymbol{K}^{-1}_{i} \right)                      \right\} = 0, \qquad i \in[1:L].
\end{align}
This proves that \eqref{ex_main3} is lower bounded by 0.

\section{Proof of Theorem \ref{main_thm}}\label{MT}

The proof of Theorem \ref{main_thm} is divided into three steps. We first adapt the argument in \cite{SM04,TD07} to show that every rate tuple in $\mathcal{R}( \boldsymbol{D}_{i}, i \in [1:L])$ is achievable, i.e., $\mathcal{R}( \boldsymbol{D}_{i}, i \in [1:L])\subseteq\mathcal{R}^*( \boldsymbol{D}_{i}, i \in [1:L])$. We then study the supporting hyperplanes of $\mathcal{R}( \boldsymbol{D}_{i}, i \in [1:L])$ and characterize the optimal solution of the relevant minimization problem via KKT analysis. Finally we derive a matching converse by leveraging the extremal inequality in Theorem \ref{ei_thm}.

%In this section, we complete the rate-distortion region characterization of the vector Gaussian successive refinement source coding system with degraded side information. We first show the admissible scheme by considering Gaussian quantization for the single-letter characterization of the discrete memoryless case in Theorem \ref{thm:TD07}. We then evaluate this admissible region by a weighted sum rate minimization problem. Finally, together with KKT analysis and the extremal inequality in Theorem \ref{ei_thm}, we prove the converse.

\subsection{Achievability}

It is easy to adapt the achevability argument in \cite{SM04,TD07} to prove the following result.
\begin{lemma}\label{lem:achievability}\it
	$(R_{i}, i \in [1:L])\in\mathcal{R}^*( \boldsymbol{D}_{i}, i \in [1:L])$ if there exist auxiliary  random vectors $(W_i, i\in[1:L])$ jointly Gaussian with $(X, Y_i, i\in[1:L])$ satisfying
	\begin{itemize}
		\item the Markov chain constraint
		\begin{align}
		(W_{i}, i \in [1:L]) \rightarrow X\rightarrow Y_{L} \rightarrow Y_{L-1} \rightarrow \ldots \rightarrow Y_{1},
		\end{align}
		\item the rate constraints
		\begin{align}
		R_{1} &\geq I (X; W_{1}| Y_{1}), \\
		\sum_{j=1}^{i} R_{j} & \geq I(X;W_{1}|Y_{1}) + \sum_{j=2}^{i}I(X;W_{j}|W_{j-1},\ldots, W_{1}, Y_{j}), \qquad  i\in[2:L],
		\end{align}		
		\item the covariance distortion constraints
		\begin{align}
		\cov(X|Y_{i}, W_{j}, j \in [1:i]) \preceq \boldsymbol{D}_{i}, \qquad i \in [1:L].
		\end{align}
	\end{itemize}	
\end{lemma}

Equipped with Lemma \ref{lem:achievability}, we proceed to show that every rate tuple in $\mathcal{R}( \boldsymbol{D}_{i}, i \in [1:L])$ is achievable. First choose auxiliary Gaussian random vectors $\left( W_{i}, i \in [1:L] \right)$ such that
\begin{align}
&\cov (X | W_{j}, j \in [1:i] ) = \left( \boldsymbol{K}_{0}^{-1}+ \sum_{j=1}^{i}\boldsymbol{B}_{j}    \right)^{-1},\qquad i\in[1:L]. \label{eq:u_con}
\end{align}
It can be verified that
\begin{align}
& h(X|Y_{i}, W_{j}, j \in [1:i]) \nonumber \\
& =  \frac{1}{2} \log \left|  (2 \pi e) \left(          \boldsymbol{K}_{0}^{-1}+ \boldsymbol{K}_{i}^{-1}  + \sum_{j=1}^{i}\boldsymbol{B}_{j}     \right)^{-1}       \right|, \qquad i\in [1:L],\label{eq:exp1}\\
& h(X|Y_{i+1}, W_{j}, j \in [1:i]) \nonumber \\
&  =  \frac{1}{2} \log \left|  (2 \pi e) \left(\boldsymbol{K}_{0}^{-1}+ \boldsymbol{K}_{i+1}^{-1}  + \sum_{j=1}^{i}\boldsymbol{B}_{j} \right)^{-1}       \right|, \quad i\in [1:L-1].\label{eq:exp2}
\end{align}
Moreover, we have
\begin{align}
&h(X|Y_{i})= h(X|X+N_{i}) =  \frac{1}{2} \log \left|  (2 \pi e) \left(          \boldsymbol{K}_{0}^{-1}+ \boldsymbol{K}_{i}^{-1}       \right)^{-1}       \right|, \qquad i\in [1:L],
\end{align}
\begin{align}
& \cov \left(    X | Y_{i},W_{j}, j \in [1:i]  \right) =   \left(          \boldsymbol{K}_{0}^{-1}+ \boldsymbol{K}_{i}^{-1}  + \sum_{j=1}^{i}\boldsymbol{B}_{j}     \right)^{-1}, \qquad i\in [1:L].
\end{align}
Now one can readily prove $\mathcal{R}(\mathbf{D}_{i}, i \in [1:L])\subseteq\mathcal{R}^*(\mathbf{D}_{i}, i \in [1:L])$ by invoking Lemma \ref{lem:achievability} and a timesharing argument.

%Thereby, we obtain an inner bound of the rate-distortion region~$\mathcal{R}(\mathbf{D}_{i}, i \in [1:L])$ by involving all the rate tuples satisfying
%\begin{align}
%R_{1} & \geq \frac{1}{2} \log \frac{| \boldsymbol{K}_{0}^{-1} + \boldsymbol{K}_{1}^{-1} + \boldsymbol{B}_{1} |}{|\boldsymbol{K}_{0}^{-1}+\boldsymbol{K}_{1}^{-1} |}, \\
%\sum_{j=1}^{i} R_{j} & \geq \frac{1}{2} \log \frac{| \boldsymbol{K}_{0}^{-1} + \boldsymbol{K}_{1}^{-1} + \boldsymbol{B}_{1} |}{|\boldsymbol{K}_{0}^{-1}+\boldsymbol{K}_{1}^{-1} |}+ \sum_{j=2}^{i}\frac{1}{2} \log \frac{| \boldsymbol{K}_{0}^{-1} + \boldsymbol{K}_{j}^{-1} + \sum_{k=1}^{j}\boldsymbol{B}_{k}  |}{|\boldsymbol{K}_{0}^{-1}+\boldsymbol{K}_{j}^{-1}+\sum_{k=1}^{j-1}\boldsymbol{B}_{k} |}, \qquad  i\in[2:L],
%\end{align}
%for sequence of symmetric and positive semi-definite matrices $(\boldsymbol{B}_{i}, i \in [1:L])$ such that
%\begin{align}
%\boldsymbol{B}_{i} &\succeq \boldsymbol{0}, \qquad i \in[1;L], \\
%\sum_{j=1}^{i} \boldsymbol{B}_{j} & \succeq \boldsymbol{D}_{i}^{-1} - \boldsymbol{K}_{0}^{-1} - \boldsymbol{K}_{i}^{-1}, \qquad i \in[1;L].
%\end{align}
%This proves the achievability part of Theorem \ref{main_thm}.

\subsection{Supporting Hyperplane Characterization}

Since  $\mathcal{R}(\mathbf{D}_{i}, i \in [1:L])$ is convex, it is completely specified by its supporting hyperplanes.
The characterization of the supporting hyperplanes boils down to solving the following optimization problem
\begin{align}
R^{*} \triangleq\inf_{(R_{1},\ldots,R_{L}) \in \mathcal{R}(\mathbf{D}_{i}, i \in [1:L])} \sum_{i=1}^{L} \mu_{i}R_{i}, \label{PG}
\end{align}
where $\mu_{1} \geq \mu_{2} \geq \ldots \geq \mu_{L}\geq 0$. It is clear that
\begin{align}
 R^{*} =&  \min_{(\mathbf{B}_{i}, i \in [1:L])}        \frac{\mu_{1}}{2} \log \frac{| \boldsymbol{K}_{0}^{-1} + \boldsymbol{K}_{1}^{-1} + \boldsymbol{B}_{1} |}{|\boldsymbol{K}_{0}^{-1}+\boldsymbol{K}_{1}^{-1} |}+            \sum_{i=2}^{L}  \frac{\mu_{i}}{2}  \log \frac{| \boldsymbol{K}_{0}^{-1} + \boldsymbol{K}_{i}^{-1} + \sum_{j=1}^{i}\boldsymbol{B}_{j}  |}{|\boldsymbol{K}_{0}^{-1}+\boldsymbol{K}_{i}^{-1}+\sum_{j=1}^{i-1}\boldsymbol{B}_{j} |}  \label{PVG} \\
{} & \mbox{subject to    } \qquad \boldsymbol{B}_{i} \succeq \boldsymbol{0}, \hspace{3.4cm} i \in[1;L],  \nonumber \\
  {} & \qquad  \qquad \; \sum_{j=1}^{i} \boldsymbol{B}_{j}  \succeq \boldsymbol{D}_{i}^{-1} - \boldsymbol{K}_{0}^{-1} - \boldsymbol{K}_{i}^{-1}, \qquad i \in[1;L]. \nonumber
\end{align}

%The necessary KKT conditions of \eqref{PVG} are given in the following theorem,

\begin{theorem} \label{theorem:KKT}
The minimizer $(\mathbf{B}^{*}_{i}, i \in [1:L])$ of \eqref{PVG} must satisfy
\begin{align}
\frac{\mu_{i}}{2} \left(    \boldsymbol{K}_{0}^{-1} + \boldsymbol{K}_{i}^{-1} + \sum_{j=1}^{i} \boldsymbol{B}_{j}^{*}          \right)^{-1}- \frac{\mu_{i+1}}{2} \left(    \boldsymbol{K}_{0}^{-1} + \boldsymbol{K}_{i+1}^{-1} + \sum_{j=1}^{i} \boldsymbol{B}_{j}^{*}          \right)^{-1} &= \boldsymbol{\Psi}_{i}-\boldsymbol{\Psi}_{i+1}+\boldsymbol{\Lambda}_{i}, \qquad i \in [1:L-1], \label{eq_KKT1}\\
\frac{\mu_{L}}{2} \left(    \boldsymbol{K}_{0}^{-1} + \boldsymbol{K}_{L}^{-1} + \sum_{j=1}^{L} \boldsymbol{B}_{j}^{*}          \right)^{-1} &=\boldsymbol{\Psi}_{L}+\boldsymbol{\Lambda}_{L}, \label{eq_KKT2}
\end{align}
for some positive semi-definite matrices $(\boldsymbol{\Psi}_{i}, i \in [1:L])$ and $(\boldsymbol{\Lambda}_{i}, i \in [1:L])$  such that
\begin{align}
\boldsymbol{B}_{i}^{*} \boldsymbol{\Psi}_{i} &= \boldsymbol{0}, \qquad i \in [1:L], \label{eq_IQ}\\
\left( \boldsymbol{K}_{0}^{-1} + \boldsymbol{K}_{i}^{-1} + \sum_{j=1}^{i} \boldsymbol{B}_{j}^{*} - \boldsymbol{D}_{i}^{-1} \right) \boldsymbol{\Lambda}_{i} &= \boldsymbol{0}, \qquad i \in [1:L]. \label{eq_eq}
\end{align}
\end{theorem}

\begin{IEEEproof}
The Lagrangian of \eqref{PVG} is given by
\begin{align}
&\frac{\mu_{1}}{2} \log \frac{| \boldsymbol{K}_{0}^{-1} + \boldsymbol{K}_{1}^{-1} + \boldsymbol{B}_{1} |}{|\boldsymbol{K}_{0}^{-1}+\boldsymbol{K}_{1}^{-1} |}+            \sum_{i=2}^{L}  \frac{\mu_{i}}{2}  \log \frac{| \boldsymbol{K}_{0}^{-1} + \boldsymbol{K}_{i}^{-1} + \sum_{j=1}^{i}\boldsymbol{B}_{j}  |}{|\boldsymbol{K}_{0}^{-1}+\boldsymbol{K}_{i}^{-1}+\sum_{j=1}^{i-1}\boldsymbol{B}_{j} |} \nonumber \\
&- \sum_{i=1}^{L} \tr\{\boldsymbol{B}_{i}\boldsymbol{\Psi}_{i}+(\boldsymbol{K}_{0}^{-1} +  \boldsymbol{K}_{i}^{-1} - \boldsymbol{D}_{i}^{-1}+ \sum_{j=1}^{i}\boldsymbol{B}_{j})\boldsymbol{\Lambda}_{i}\},
\end{align}
where positive semi-definite matrices $(\boldsymbol{\Psi}_{i}, i \in [1:L])$ and $(\boldsymbol{\Lambda}_{i}, i \in [1:L])$ serve as Lagrange multipliers. Note that \eqref{eq_KKT1}-\eqref{eq_eq} follow directly form the KKT  conditions. The proof is complete by verifying a set of constraint qualifications in \cite[Sections 4-5]{book}.
\end{IEEEproof}

\begin{remark}
	It is worth noting that \eqref{eq_KKT1}-\eqref{eq_eq} in Theorem \ref{theorem:KKT} correspond exactly to \eqref{id_KKT1}-\eqref{CS_eq} in Theorem \ref{ei_thm}.
\end{remark}

\subsection{Converse}

It is easy to adapt the converse argument in
\cite{SM04,TD07} to prove the following result.
\begin{lemma}\label{lem:converse}\it
	For any $(R_{i}, i \in [1:L])\in\mathcal{R}^*( \boldsymbol{D}_{i}, i \in [1:L])$ and any $\epsilon>0$, there exist auxiliary random objects jointly distributed with $(X, Y_i, i\in[1:L])$ satisfying
	\begin{itemize}
		\item the Markov chain constraint
		\begin{align}
		(W_{i}, i \in [1:L]) \rightarrow X\rightarrow Y_{L} \rightarrow Y_{L-1} \rightarrow \ldots \rightarrow Y_{1},
		\end{align}
		\item the rate constraints
		\begin{align}
		R_{1}+\epsilon &\geq I (X; W_{1}| Y_{1}), \\
		\sum_{j=1}^{i} (R_{j}+\epsilon)& \geq I(X;W_{1}|Y_{1}) + \sum_{j=2}^{i}I(X;W_{j}|W_{j-1},\ldots, W_{1}, Y_{j}), \qquad  i\in[2:L],
		\end{align}		
		\item the covariance distortion constraints
		\begin{align}
		\cov(X|Y_{i}, W_{j}, j \in [1:i]) \preceq \boldsymbol{D}_{i}+\epsilon\mathbf{I}, \qquad i \in [1:L].
		\end{align}
	\end{itemize}
\end{lemma}

 Now we proceed to show that $\mathcal{R}^*( \boldsymbol{D}_{i}, i \in [1:L])\subseteq\mathcal{R}( \boldsymbol{D}_{i}, i \in [1:L])$. For any $(R_{1},\ldots,R_{L}) \in \mathcal{R}^*(\mathbf{D}_{i}, i \in [1:L])$ and any $\epsilon>0$, it follows by Lemma \ref{lem:converse}, Theorem \ref{theorem:KKT}, and Theorem \ref{ei_thm} that
\begin{align}
& \sum_{i=1}^{L} \mu_{i}(R_{i}+\epsilon)\nonumber\\
&\geq \mu_{1} I(X;W_{1}|Y_{1}) + \sum_{j=2}^{i}\mu_{i}I(X;W_{j}|W_{j-1},\ldots, W_{1}, Y_{j}) \label{eq:126}\\
&=\mu_{1}h(X|Y_{1}) +\sum_{i=1}^{L-1}    \left(    \mu_{i}h(Y_{i}| W_{j}, j \in [1:i] ) - \mu_{i+1}h(Y_{i+1}| W_{j}, j \in [1:i]) - (\mu_{i} - \mu_{i+1})h(X | W_{j}, j \in [1:i])                     \right)      \nonumber  \\
&\quad + \mu_{L} h(Y_{L}| W_{j}, j \in [1:L]) - \mu_{L} h(X | W_{j}, j \in [1:L])  \\
& \geq -\frac{\mu_{1}}{2} \log \left|  (2 \pi e)^{-1} \left(          \boldsymbol{K}_{0}^{-1}+ \boldsymbol{K}_{i}^{-1}       \right)       \right|+ \sum_{i=1}^{L-1}  \left(  -\frac{\mu_{i+1}}{2} \log \left| (2 \pi e)^{-1}\left( \boldsymbol{K}_{0}^{-1} + \boldsymbol{K}_{i+1}^{-1} + \sum_{j=1}^{i} \boldsymbol{B}_{j}^{*}(\epsilon)   \right)  \right| \right.\nonumber \\
& \quad \left. + \frac{\mu_{i}}{2} \log \left| (2 \pi e)^{-1}\left(    \boldsymbol{K}_{0}^{-1} + \boldsymbol{K}_{i}^{-1} + \sum_{j=1}^{i} \boldsymbol{B}_{j}^{*}  (\epsilon)        \right)  \right| \right) +  \frac{\mu_{L}}{2} \log \left| (2 \pi e)^{-1} \left(    \boldsymbol{K}_{0}^{-1} + \boldsymbol{K}_{L}^{-1} + \sum_{j=1}^{L} \boldsymbol{B}_{j}^{*}(\epsilon)          \right)  \right|, \label{eq:128}
\end{align}
where $(\mathbf{B}^{*}_{i}(\epsilon), i \in [1:L])$ denotes the minimizer of \eqref{PVG} with $(\mathbf{D}_{i}, i \in [1:L])$ replaced by  $(\mathbf{D}_{i}+\epsilon\mathbf{I}, i \in [1:L])$. Now one can readily show
\begin{align}
\sum_{i=1}^{L} \mu_{i}R_{i}\geq R^*
\end{align}
via a simple limiting argument.
This completes the proof of Theorem \ref{main_thm}.

%m the lower bound for discrete memoryless sources in Theorem \ref{thm:TD07}, \eqref{eq:128} is from the extremal inequality in Theorem \ref{ei_thm}. Since the lower bound is admissible with Gaussian quantization, this concludes the proof of Theorem \ref{main_thm}.

\section{Conclusion}\label{sec:conclusion}

We have studied the problem of successive refinement for  Wyner-Ziv coding with degraded side information and obtained a computable characterization of the rate region in the quadratic vector Gaussian setting. From the technical perspective, our main contribution is a new extremal inequality, which is established via a refined monotone path argument inspired by the doubling trick in \cite{GN14}. In a recent paper  \cite{UW17},
Unal and Wagner considered the vector Gaussian Heegard-Berger/Kaspi problem with no degradedness assumption on side information and obtained several conclusive results through careful comparisons of the relevant covariance distortions. In contrast, our proof technique does not require such comparisons and thus is potentially better suited to  the non-degraded side information case. It is of considerable interest to investigate whether this technique can yield new results beyond those in \cite{UW17}.

% Therefore, it is of considerable interest to investigate the application of this proof technique to the non-degraded side information case

%In this paper, we consider the vector Gaussian successive refinement source coding problem with degraded side information, and fully characterize the rate-distortion region under the covariance mean square error distortion. The characterization is based on a new extremal inequality. The proof of this extremal inequality relies on the combination Gaussian perturbation construction in \cite{WC19}, and information-estimation relationship. One recent work we should mention is \cite{UW17}, in which Unal and Wagner considered the vector Gaussian Heegard-Berger/Kaspi problem without such a degraded assumption on side information. An outer bound was obtained by using enhancement argument. With careful comparisons of different covariance matrix distortions, they obtained several interesting instances matching their outer bound. In this paper, since our MMSE evaluation method avoid the comparisons of the covariance matrices, it may be helpful to investigate the non-degraded case as in \cite{UW17}.

\appendices

\section{Preliminaries on Fisher Information and MMSE} \label{app_lea2}

Here is a summary of some basic properties of Fisher information and MMSE, which will be used extensively in the proof of extremal inequality \eqref{ex_inq}.

%As the proof of extremal inequality \eqref{ex_inq} relies heavily on Fisher information matrices and MMSE (minimum mean square error) matrices, we shall review some basic properties of Fisher information and MMSE before going into the formal proof steps.

We begin with the definition of conditional Fisher information matrix and MMSE matrix.
\smallskip
\begin{definition}\it
Let $({X}, U)$ be a pair of jointly distributed random vectors with differentiable conditional probability density function:
\begin{equation}
f(\boldsymbol{x}|u) \triangleq f(x_{i}, i \in [1:m]|u).
\end{equation}
The vector-valued score function is defined as
\begin{equation}
\nabla \log f(\boldsymbol{x}| u)  = \left[\frac{\partial \log f(\boldsymbol{x}| u)}{\partial x_{1}}, \cdots, \frac{\partial \log f(\boldsymbol{x}| u)}{\partial x_{m}} \right]^{T}.
\end{equation}
The conditional Fisher information of $X$ respect to $U$ is given by
\begin{equation}
J(X| U) = \mathbb{E}\left[  \left(\nabla \log f(\boldsymbol{x}| u) \right) \cdot \left(\nabla \log f(\boldsymbol{x} | u) \right)^{T}    \right ].
\end{equation}
\end{definition}

\smallskip
\begin{definition}\label{def_MMSE}\it
Let $(X,Y,U)$ be a set of jointly distributed random vectors. The conditional covariance matrix of $X$ given $(Y,U)$ is defined as
\begin{equation}
\cov (X|Y,U) = \mathbb{E}\left[    \left( X - \mathbb{E}[X|Y,U] \right)  \cdot    \left( X - \mathbb{E}[X|Y,U] \right)^{T}           \right].
\end{equation}
\end{definition}

\smallskip
\begin{lemma} [Matrix Version of de Bruijn's Identity]\label{de}\it
Let $({X}, U)$ be a pair of jointly distributed random vectors, and ${N} \thicksim {N}(\mathbf{0}, \boldsymbol{\Sigma})$ be a Gaussian random vector independent of  $({X}, U)$. Then
 \begin{equation}
 \nabla_{\boldsymbol{\Sigma} }h({X}+  {N} | U) = \frac{1}{2}  J ({X}+{N} | U) . \label{eq:de}
 \end{equation}
\end{lemma}

Lemma \ref{de} is a conditional version of \cite[Theorem 1]{PV06}, which provides a link between differential entropy and Fisher information.
\smallskip
\begin{lemma}\label{lea1}\it
Let $(X, U)$ be a pair of jointly distributed random vectors, and $N \thicksim \mathcal{N}(\boldsymbol{0}, \boldsymbol{\Sigma})$ be a Gaussian random vector independent of  $(X, U)$.  Then
\begin{equation}
 J(X + N | U) + \boldsymbol{\Sigma}^{-1} \cov( X | X +  N, U) \boldsymbol{\Sigma}^{-1} = \boldsymbol{\Sigma}^{-1}. \label{eq_lea1}
 \end{equation}
\end{lemma}

The complementary identity in Lemma \ref{lea1} provides a link between Fisher information and MMSE, and its proof can be found in \cite[Corollary 1]{PV06}.
\smallskip

\smallskip
\begin{lemma} \label{fi_inq}\it
Let $({X}, {Y},U)$ be a set of jointly distributed random vectors. Assume that ${X}$ and ${Y}$ are
conditionally independent given $U$.  Then for any square matrix $\boldsymbol{A}$ and $\boldsymbol{B}$,
\begin{equation}
(\boldsymbol{A}+ \boldsymbol{B})J({X}+ {Y} | U)(\boldsymbol{A}+ \boldsymbol{B})^{T} \preceq \boldsymbol{A} J({X} | U)\boldsymbol{A}^{T} + \boldsymbol{B} J({Y} | U)\boldsymbol{B}^{T}. \label{inq:bod}
\end{equation}
\end{lemma}
\begin{IEEEproof}
From the conditional version of matrix Fisher information inequality in \cite[Appendix II]{LV07}, we have
\begin{equation}
J(X+Y|U) \preceq \boldsymbol{K}J(X|U)\boldsymbol{K}^{T} + (\boldsymbol{I} - \boldsymbol{K})J(Y|U)(\boldsymbol{I} - \boldsymbol{K})^{T},
\end{equation}
for any square matrix $\boldsymbol{K}$. Setting
\begin{equation}
\boldsymbol{K} = (\boldsymbol{A}+ \boldsymbol{B})^{-1}\boldsymbol{A}
\end{equation}
proves  \eqref{inq:bod}.
\end{IEEEproof}

\smallskip

\begin{lemma} \label{MMSE_inq}\it
Let $X$ be a Gaussian random vector and $U$ be an arbitrary random vector. Let $N_{1}$ and $N_{2}$ be two zero-mean Gaussian random vectors, independent of $(X,U)$, with covariance matrices $\boldsymbol{\Sigma}_{1}$ and $\boldsymbol{\Sigma}_{2}$, respectively. If
\begin{equation}
\boldsymbol{\Sigma}_{2} \succ \boldsymbol{\Sigma}_{1} \succ \boldsymbol{0},
\end{equation}
then
\begin{equation}
\cov \left( X \big| X+N_{1},U\right)^{-1} - \boldsymbol{\Sigma}_{1}^{-1} \succeq \cov \left( X \big| X+N_{2},U\right)^{-1} - \boldsymbol{\Sigma}_{2}^{-1}.
\end{equation}
\end{lemma}

Lemma \ref{MMSE_inq} can be proved by combining the Cram\'{e}r-Rao inequality and the complementary identity in Lemma \ref{lea1}. See \cite[Lemma 4]{WC19} for details.

%The proof is followed by combing Cram\'{e}r-Rao inequality and complementary identity in Lemma \ref{lea1}. One can refer proof details in \cite[Lemma 4]{WC19}.

\smallskip
\begin{lemma} [Data Processing Inequality for Fisher Information] \label{DP_FI}\it
Let $({X}, U, V)$ be a set of jointly distributed random vectors. Assume that $U \rightarrow V \rightarrow {X}$ form a Markov chain. Then
\begin{equation}
J({X} | U) \preceq J({X} | V).
\end{equation}
\end{lemma}

Lemma \ref{DP_FI} is analogous to \cite[Lemma 3]{Z98}, and can be easily proved using the chain rule of Fisher information matrix \cite[Lemma 1]{Z98}.

%The proof is easily followed by the chain rule of Fisher information matrix \cite[Lemma 1]{Z98}, which is analogous to another form of data processing inequality \cite[Lemma 3]{Z98}.

\smallskip
\begin{lemma} [Data Processing Inequality for MMSE] \label{DP_MMSE}\it
Let $({X}, U, V)$ be a set of jointly distributed random vectors. Assume $U \rightarrow V \rightarrow {X}$ form a Markov chain. Then
\begin{equation}
\cov ({X} | U) \succeq \cov ({X} | V).
\end{equation}
\end{lemma}

See \cite[Proposition 5]{R11} for a detailed proof of Lemma \ref{DP_MMSE}.

\smallskip

\section{Derivative of the bivariate differential entropy $h\left({X}_{i, \gamma}, {Y}^{*}_{i, \gamma}| W_{j}, j \in [1:i]\right)$} \label{app_tev1}

In view of \eqref{eqn:per1} and \eqref{eqn:per5}, we have
\begin{align}
&h\left({X}_{i, \gamma}, {Y}^{*}_{i, \gamma} | W_{j}, j \in [1:i]\right)\\
&=h\left(       \sqrt{1-\gamma}{X} +  \sqrt{\gamma} {X}_{i}^{G}, \sqrt{\gamma}Y_{i}- \sqrt{1-\gamma}{{Y}}_{i}^{G}   | W_{j}, j \in [1:i]  \right)\\
&=h\left(        X + \sqrt{\frac{\gamma}{1-\gamma}} {X}_{i}^{G}, Y_{i} - \sqrt{\frac{1-\gamma}{\gamma}}    {{Y}}_{i}^{G}   \Big| W_{j}, j \in [1:i]                        \right)+\frac{n}{2} \log \gamma + \frac{n}{2} \log (1-\gamma).
\end{align}
Recall from \eqref{eq:TYG} that
\begin{equation}
{Y}_{i}^{G} = X_{i}^{G} + N_{i}^{G}.
\end{equation}
The covariance matrix of
$$
\left(\begin{array}{r}
 \sqrt{{\gamma}/{(1-\gamma)}} {X}_{i}^{G}\\
-\sqrt{{(1-\gamma)}/{\gamma}}    {{Y}}_{i}^{G}
\end{array}\right)
$$
is given by
\begin{equation}\label{eq:cov}
\boldsymbol{\Sigma}_{i,*} \triangleq \begin{pmatrix}
\frac{\gamma}{1-\gamma}\boldsymbol{\Delta}_{i} & -\boldsymbol{\Delta}_{i}\\
-\boldsymbol{\Delta}_{i} & \frac{1-\gamma}{\gamma} (  \boldsymbol{\Delta}_{i} + \boldsymbol{K}_{i}   )
\end{pmatrix}.
\end{equation}
It is easy to verify that
\begin{equation}\label{eq:covi}
\boldsymbol{\Sigma}_{i,*}^{-1} =
 \begin{pmatrix}
\frac{1-\gamma}{\gamma}( \boldsymbol{\Delta}_{i}^{-1} + \boldsymbol{K}_{i}^{-1}) & \boldsymbol{K}_{i}^{-1}\\
\boldsymbol{K}_{i}^{-1} & \frac{\gamma}{1-\gamma}  \boldsymbol{K}_{i}^{-1}
\end{pmatrix}
\end{equation}
and
\begin{equation}\label{eq:covd}
\nabla_{\gamma}{\boldsymbol{\Sigma}_{i,*} } =  \begin{pmatrix}
\frac{1}{(1-\gamma)^{2}}\boldsymbol{\Delta}_{i} & \boldsymbol{0}\\
\boldsymbol{0} & -\frac{1}{\gamma^{2}} (  \boldsymbol{\Delta}_{i} + \boldsymbol{K}_{i}   )
\end{pmatrix}.
\end{equation}
Combining \eqref{eq:covi} and \eqref{eq:covd} gives
\begin{equation}\label{eq:cov1}
\tr \left\{ \left(\nabla_{\gamma}{\boldsymbol{\Sigma}_{i,*} }\right) \boldsymbol{\Sigma}_{i,*}^{-1}  \right\}=0,
\end{equation}
\begin{equation}\label{eq:cov2}
\boldsymbol{\Sigma}_{i,*}^{-1}\left( \nabla_{\gamma}{\boldsymbol{\Sigma}_{i,*} } \right) \boldsymbol{\Sigma}_{i,*}^{-1} = \begin{pmatrix}
-\frac{1}{\gamma^{2}}(\boldsymbol{\Delta}_{i}^{-1}+ \boldsymbol{K}_{i}^{-1} )& \boldsymbol{0}\\
\boldsymbol{0} & \frac{1}{(1-\gamma)^{2}}  \boldsymbol{K}_{i}^{-1}
\end{pmatrix}.
\end{equation}

By invoking the chain rule of matrix calculus and Lemma \ref{de} in Appendix \ref{app_lea2}, we have
\begin{align}
&\frac{d}{d\gamma}h\left({X}_{i, \gamma}, {Y}^{*}_{i, \gamma} | W_{j}, j \in [1:i]\right) \nonumber \\
&= \frac{d}{d\gamma} \left\{ h\left(        X + \sqrt{\frac{\gamma}{1-\gamma}} {X}_{i}^{G}, Y_{i} - \sqrt{\frac{1-\gamma}{\gamma}}    {{Y}}_{i}^{G}   \Big| W_{j}, j \in [1:i]                        \right)+\frac{n}{2} \log \gamma + \frac{n}{2} \log (1-\gamma) \right\} \\
&= \frac{1}{2} \tr \left\{\left( \nabla_{\gamma}{\boldsymbol{\Sigma}_{i,*} } \right)  J \left(
\begin{pmatrix}
\sqrt{\frac{1}{1-\gamma}}{{X}^{T}_{i, \gamma}} & \sqrt{\frac{1}{\gamma}}{{Y}^{*}_{i, \gamma}}^{T}
\end{pmatrix}^{T}
\Big| W_{j}, j \in [1:i]                        \right)            \right\} + \frac{n}{2} \left(  \frac{1}{\gamma} - \frac{1}{1-\gamma} \right).\label{eq:cov5}
\end{align}
It can be verified
\begin{align}
& \tr \left\{ \left(\nabla_{\gamma}{\boldsymbol{\Sigma}_{i,*} } \right)  J \left(
\begin{pmatrix}
\sqrt{\frac{1}{1-\gamma}}{{X}^{T}_{i, \gamma}} & \sqrt{\frac{1}{\gamma}}{{Y}^{*}_{i, \gamma}}^{T}
\end{pmatrix}^{T}
\Big| W_{j}, j \in [1:i]                        \right)           \right\}\nonumber \\
& = \tr  \Big\{ \left(\nabla_{\gamma}{\boldsymbol{\Sigma}_{i,*} }\right) \boldsymbol{\Sigma}_{i,*}^{-1}  -    \boldsymbol{\Sigma}_{i,*}^{-1} \left( \nabla_{\gamma}{\boldsymbol{\Sigma}_{i,*} } \right) \boldsymbol{\Sigma}_{i,*}^{-1}\nonumber \\
&\qquad \qquad  \cov \left(
 \begin{pmatrix}
X^{T} & Y_{i}^{T}
\end{pmatrix}^{T}
 \Big|   X + \sqrt{\frac{\gamma}{1-\gamma}} {X}_{i}^{G}, Y_{i} - \sqrt{\frac{1-\gamma}{\gamma}}    {{Y}}_{i}^{G}, W_{j}, j \in [1:i]     \right)       \Big\} \label{eq:compli}\\
&= \tr \Big\{           \begin{pmatrix}
-\frac{1}{\gamma^{2}}(\boldsymbol{\Delta}_{i}^{-1}+ \boldsymbol{K}_{i}^{-1} )& \boldsymbol{0}\\
\boldsymbol{0} & \frac{1}{(1-\gamma)^{2}}  \boldsymbol{K}_{i}^{-1}
\end{pmatrix} \nonumber\\
&
\qquad \qquad \cov \left(
 \begin{pmatrix}
X^{T} & Y_{i}^{T}
\end{pmatrix}^{T}
\Big|   X + \sqrt{\frac{\gamma}{1-\gamma}} {X}_{i}^{G}, Y_{i} - \sqrt{\frac{1-\gamma}{\gamma}}    {{Y}}_{i}^{G}, W_{j}, j \in [1:i]     \right)                    \Big\}, \label{eq:cov3}
\end{align}
where  \eqref{eq:compli} follows by Lemma \ref{lea1} in Appendix \ref{app_lea2}, and \eqref{eq:cov3} is due to \eqref{eq:cov1} and \eqref{eq:cov2}.
Notice that
\begin{align}
& \cov \left(
 \begin{pmatrix}
X^{T} & Y_{i}^{T}
\end{pmatrix}^{T}
\Big|    X + \sqrt{\frac{\gamma}{1-\gamma}} {X}_{i}^{G}, Y_{i} - \sqrt{\frac{1-\gamma}{\gamma}}    {{Y}}_{i}^{G}   \right)   \nonumber \\
& =
\left(
\begin{pmatrix}
\boldsymbol{K}_{0} & \boldsymbol{K}_{0} \\
\boldsymbol{K}_{0} & \boldsymbol{K}_{0}+ \boldsymbol{K}_{i}
\end{pmatrix}^{-1}
+ \boldsymbol{\Sigma}_{i,*}^{-1}
\right)^{-1} \\
& =
\left(
\begin{pmatrix}
\boldsymbol{K}_{0}^{-1}+\boldsymbol{K}_{i}^{-1} & -\boldsymbol{K}_{i}^{-1} \\
-\boldsymbol{K}_{i}^{-1} & \boldsymbol{K}_{i}^{-1}
\end{pmatrix}
+
\begin{pmatrix}
\frac{1-\gamma}{\gamma}( \boldsymbol{\Delta}_{i}^{-1} + \boldsymbol{K}_{i}^{-1}) & \boldsymbol{K}_{i}^{-1}\\
\boldsymbol{K}_{i}^{-1} & \frac{\gamma}{1-\gamma}  \boldsymbol{K}_{i}^{-1}
\end{pmatrix}
\right)^{-1} \\
& =
\begin{pmatrix}
\left(  \boldsymbol{K}_{0}^{-1}+  \frac{1-\gamma}{\gamma} \boldsymbol{\Delta}_{i}^{-1} + \frac{1}{\gamma}\boldsymbol{K}_{i}^{-1}      \right)^{-1} & \boldsymbol{0}\\
\boldsymbol{0} & (1-\gamma)  \boldsymbol{K}_{i}
\end{pmatrix}.
\end{align}
Thus, we have the Markov chain
\begin{equation}
 \left( W_{j}, j \in [1:i] \right) \rightarrow X \rightarrow \left(  X + \sqrt{\frac{\gamma}{1-\gamma}} {X}_{i}^{G}, Y_{i} - \sqrt{\frac{1-\gamma}{\gamma}}    {{Y}}_{i}^{G}     \right) \rightarrow Y_{i}.
\end{equation}
As a consequence,
\begin{align}
& \cov \left(
 \begin{pmatrix}
 X^{T} & Y_{i}^{T}
\end{pmatrix}^{T}
 \Big|    {X}_{i, \gamma}, {Y}^{*}_{i, \gamma},  W_{j}, j \in [1:i]     \right)     \nonumber \\
& =
\begin{pmatrix}
\cov \left(   X \Big | {X}_{i, \gamma}, {Y}_{i, \gamma}, W_{j}, j \in [1:i]     \right) & \boldsymbol{0}\\
\boldsymbol{0} &  (1-\gamma)  \boldsymbol{K}_{i} \label{eq:cov4}
\end{pmatrix}.
\end{align}
%
%On the other hand, by invoking the generalized complementary identity in Lemma \ref{lea2}, we have the following Fisher information representation of MMSE matrix in \eqref{eq:cov4},
%\begin{align}
%&\cov \left(   X \Big |  X - \sqrt{\frac{1-\gamma}{\gamma}} {X}_{i}^{G}, Y_{i+1} + \sqrt{\frac{\gamma}{1-\gamma}}    {\tilde{Y}}_{i+1}^{G}, W_{j}, j \in [1:i]     \right) \nonumber \\
%&=\cov \left(   X \Big |  X - \sqrt{\frac{1-\gamma}{\gamma}} {X}_{i}^{G}, X+(1-\gamma)N_{i+1} + \sqrt{\gamma (1-\gamma) }N_{i+1}^{G}, W_{j}, j \in [1:i]     \right) \\
%&= \left(   \frac{\gamma}{1-\gamma}\boldsymbol{\Delta}_{i}^{-1}   + \frac{1}{1-\gamma}\boldsymbol{K}_{i+1}^{-1}   \right)^{-1} \nonumber \\
% &\quad - J\left(     \left(   \frac{\gamma}{1-\gamma}\boldsymbol{\Delta}_{i}^{-1}   + \frac{1}{1-\gamma}\boldsymbol{K}_{i+1}^{-1}   \right)X -\sqrt{\frac{\gamma}{1-\gamma}}\boldsymbol{\Delta}_{i}^{-1}X_{i}^{G} + \boldsymbol{K}_{i+1}^{-1} N_{i+1} +   \sqrt{\frac{\gamma}{1-\gamma}}\boldsymbol{K}_{i+1}^{-1} N_{i+1}^{G}        \Big | W_{j}, j \in [1:i]   \right)
%\end{align}
By combining \eqref{eq:cov5}, \eqref{eq:cov3} and \eqref{eq:cov4}, we obtain
\begin{align}
&\frac{d}{d\gamma}h\left({X}_{i, \gamma}, {Y}^{*}_{i, \gamma}\Big| W_{j}, j \in [1:i]\right) \nonumber \\
& = \frac{1}{2} \tr \left\{
\begin{pmatrix}
-\frac{1}{\gamma^{2}}(\boldsymbol{\Delta}_{i}^{-1}+ \boldsymbol{K}_{i}^{-1} )& \boldsymbol{0}\\
\boldsymbol{0} & \frac{1}{(1-\gamma)^{2}}  \boldsymbol{K}_{i}^{-1}
\end{pmatrix}
\begin{pmatrix}
\cov \left(   X \Big | {X}_{i, \gamma}, {Y}^{*}_{i, \gamma},  W_{j}, j \in [1:i]     \right) & \boldsymbol{0}\\
\boldsymbol{0} &  (1-\gamma)  \boldsymbol{K}_{i} \label{eq:covm4}
\end{pmatrix}
\right\} \nonumber \\
& \quad  + \frac{n}{2} \left(  \frac{1}{\gamma} - \frac{1}{1-\gamma} \right)   \\
& = -\frac{1}{2\gamma} \tr \left\{   \frac{1}{\gamma} (\boldsymbol{\Delta}_{i}^{-1}+ \boldsymbol{K}_{i}^{-1} ) \cov \left(   X \Big | {X}_{i, \gamma}, {Y}^{*}_{i, \gamma},  W_{j}, j \in [1:i]     \right) \ -\boldsymbol{I}   \right\}\\
%& = -\frac{1}{2\gamma} \tr \left\{   (\boldsymbol{\Delta}_{i}^{-1}+ \boldsymbol{K}_{i}^{-1} )^{-1}  \right.\nonumber \\
%& \qquad \qquad \left. \left(    \frac{1}{\gamma} (\boldsymbol{\Delta}_{i}^{-1}+ \boldsymbol{K}_{i}^{-1} ) \cov \left(   X \Big | {X}_{i, \gamma}, {Y}^{*}_{i, \gamma},  W_{j}, j \in [1:i]     \right) (\boldsymbol{\Delta}_{i}^{-1}+ \boldsymbol{K}_{i}^{-1} ) - (\boldsymbol{\Delta}_{i}^{-1}+ \boldsymbol{K}_{i}^{-1} )              \right)      \right\}
&=-\frac{1}{2\gamma} \tr \left\{   (\boldsymbol{\Delta}_{i}^{-1}+ \boldsymbol{K}_{i}^{-1} ) \left( \frac{1}{\gamma} \cov \left(   X \Big | {X}_{i, \gamma}, {Y}^{*}_{i, \gamma},  W_{j}, j \in [1:i]     \right) \ - (\boldsymbol{\Delta}_{i}^{-1}+ \boldsymbol{K}_{i}^{-1} )^{-1} \right)  \right\}. \label{eq:tt1.2}
\end{align}

On the other hand, it follows by the theory of linear MMSE estimation that
\begin{align}
\sqrt{\gamma}X_{i}^{G} = -\sqrt{\gamma (1-\gamma)} \left(    \boldsymbol{\Delta}_{i}^{-1} + \left( 1-\gamma \right)\boldsymbol{K}_{i}^{-1}   \right)^{-1}\boldsymbol{K}_{i}^{-1}\left( \sqrt{\gamma} N_{i} -\sqrt{1-\gamma}Y_{i}^{G}      \right) + \sqrt{\gamma}\hat{X}_{i}^{G},
\end{align}
where $\hat{X}_{i,\gamma}$ is a Gaussian random vector with mean zero and covariance matrix $ \left(    \boldsymbol{\Delta}_{i}^{-1} + \left( 1-\gamma \right)\boldsymbol{K}_{i}^{-1}   \right)^{-1}$, and  is independent of $\sqrt{\gamma} N_{i} -\sqrt{1-\gamma}Y_{i}^{G}  $.
Thus, we have
\begin{align}
X_{i,\gamma} &= \sqrt{1-\gamma}X + \sqrt{\gamma}X_{i}^{G}\\
&=\sqrt{1-\gamma}X-\sqrt{\gamma (1-\gamma)} \left(    \boldsymbol{\Delta}_{i}^{-1} + \left( 1-\gamma \right)\boldsymbol{K}_{i}^{-1}   \right)^{-1}\boldsymbol{K}_{i}^{-1}\left( \sqrt{\gamma} N_{i} -\sqrt{1-\gamma}Y_{i}^{G}      \right) + \sqrt{\gamma}\hat{X}_{i}^{G}\\
&=\sqrt{1-\gamma} \left(    \boldsymbol{\Delta}_{i}^{-1} + \left( 1-\gamma \right)\boldsymbol{K}_{i}^{-1}   \right)^{-1} \left(    \boldsymbol{\Delta}_{i}^{-1} + \boldsymbol{K}_{i}^{-1}   \right)X+ \sqrt{\gamma}\hat{X}_{i}^{G}\nonumber\\
&\quad-\sqrt{\gamma (1-\gamma)} \left(    \boldsymbol{\Delta}_{i}^{-1} + \left( 1-\gamma \right)\boldsymbol{K}_{i}^{-1}   \right)^{-1}\boldsymbol{K}_{i}^{-1}Y^{*}_{i,\gamma}.
\end{align}
The complementary Fisher information representation of $\cov \left(   X \Big | {X}_{i, \gamma}, {Y}^{*}_{i, \gamma},  W_{j}, j \in [1:i]     \right)$ can thereby be expressed as
\begin{align}
&\cov \left(   X \Big | {X}_{i, \gamma}, {Y}^{*}_{i, \gamma},  W_{j}, j \in [1:i]     \right) \\
&=\cov \left(   X \Big |\sqrt{1-\gamma} \left(    \boldsymbol{\Delta}_{i}^{-1} + \left( 1-\gamma \right)\boldsymbol{K}_{i}^{-1}   \right)^{-1} \left(    \boldsymbol{\Delta}_{i}^{-1} + \boldsymbol{K}_{i}^{-1}   \right)X+ \sqrt{\gamma}\hat{X}_{i}^{G} , {Y}^{*}_{i, \gamma},  W_{j}, j \in [1:i]     \right)\\
&=\frac{\gamma}{1-\gamma}\left(    \boldsymbol{\Delta}_{i}^{-1} + \boldsymbol{K}_{i}^{-1}   \right)^{-1}\left(    \boldsymbol{\Delta}_{i}^{-1} + \left( 1-\gamma \right)\boldsymbol{K}_{i}^{-1} -\gamma J\left(  {X}_{i, \gamma}\Big |  {Y}^{*}_{i, \gamma},  W_{j}, j \in [1:i]   \right)  \right)        \left(    \boldsymbol{\Delta}_{i}^{-1} + \boldsymbol{K}_{i}^{-1}   \right)^{-1}. \label{eq:tt1}
\end{align}
Equivalently, we can write
\begin{align}
 &\left(    \boldsymbol{\Delta}_{i}^{-1} + \boldsymbol{K}_{i}^{-1}   \right)      \left( \frac{1}{\gamma} \cov \left(   X \Big | {X}_{i, \gamma}, \tilde{Y}^{*}_{i, \gamma},  W_{j}, j \in [1:L]     \right) \ - (\boldsymbol{\Delta}_{i}^{-1}+ \boldsymbol{K}_{i}^{-1} )^{-1} \right)         \left(    \boldsymbol{\Delta}_{i}^{-1} + \boldsymbol{K}_{i}^{-1}   \right) \\
 &=\frac{\gamma}{1-\gamma}\boldsymbol{\Delta}_{i}^{-1} -\frac{\gamma}{1-\gamma} J\left(  {X}_{i, \gamma}\Big |  {Y}^{*}_{i, \gamma},  W_{j}, j \in [1:i]   \right). \label{eq:tt1.1}
\end{align}
Finally, substituting \eqref{eq:tt1.1} into \eqref{eq:tt1.2} gives
\begin{align}
&\frac{d}{d\gamma}h\left({X}_{i, \gamma}, {Y}^{*}_{i, \gamma}\Big| W_{j}, j \in [1:i]\right) \nonumber \\
&=\frac{1}{2(1-\gamma)} \tr \left\{   (\boldsymbol{\Delta}_{i}^{-1}+ \boldsymbol{K}_{i}^{-1} )^{-1} \left(  J\left(  {X}_{i, \gamma}\Big |  {Y}^{*}_{i, \gamma},  W_{j}, j \in [1:i]   \right) \ - \boldsymbol{\Delta}_{i}^{-1} \right)  \right\}.
\end{align}

%It can be verified
%\begin{align}
%& \mathbb{E} \left[     X  \Bigg|    X - \sqrt{\frac{1-\gamma}{\gamma}} {X}_{i}^{G}, Y_{i+1} + \sqrt{\frac{\gamma}{1-\gamma}}    {\tilde{Y}}_{i+1}^{G}, W_{j}, j \in [1:i]    \right] \nonumber \\
%& =
%\begin{pmatrix}
%\boldsymbol{K}_{0} & \boldsymbol{K}_{0}
%\end{pmatrix}
%\left(     \begin{pmatrix}
%\boldsymbol{K}_{0} & \boldsymbol{K}_{0}\\
%\boldsymbol{K}_{0} & \boldsymbol{K}_{0}+\boldsymbol{N}_{i+1}
%\end{pmatrix}
%+ \boldsymbol{\Sigma}_{i,*}  \right)^{-1}
%\begin{pmatrix}
% X - \sqrt{\frac{1-\gamma}{\gamma}} {X}_{i}^{G}\\
%Y_{i+1} + \sqrt{\frac{\gamma}{1-\gamma}}    {\tilde{Y}}_{i+1}^{G}
%\end{pmatrix}
%\end{align}
%

\section{Derivative of the bivariate differential entropy $h\left(\tilde{Y}_{i+1, \gamma}, {Y}^{*}_{i, \gamma}| W_{j}, j \in [1:i]\right)$} \label{app_tev2}

In view of \eqref{eqn:per4} and \eqref{eqn:per5},
\begin{align}
&h\left(\tilde{Y}_{i+1, \gamma}, {Y}^{*}_{i, \gamma}| W_{j}, j \in [1:i]\right)\\
&=h\left(       \sqrt{1-\gamma}{Y}_{i+1} +  \sqrt{\gamma} \tilde{Y}_{i+1}^{G}, \sqrt{\gamma}Y_{i}- \sqrt{1-\gamma}{{Y}}_{i}^{G}   \Big| W_{j}, j \in [1:i]  \right)\\
&=h\left(    Y_{i+1} + \sqrt{\frac{\gamma}{1-\gamma}} \tilde{Y}_{i+1}^{G}, Y_{i} - \sqrt{\frac{1-\gamma}{\gamma}}    {{Y}}_{i}^{G}   \Big| W_{j}, j \in [1:i]                        \right)+\frac{n}{2} \log \gamma + \frac{n}{2} \log (1-\gamma).
\end{align}
By the definition of $ {{Y}}_{i}^{G}$ and $\tilde{Y}_{i+1}^{G}$ in \eqref{eq:TYG} and \eqref{eq:YG} as well as the construction of $  \left( N_{i}^{G}, i \in [1:L] \right) $, we can write
\begin{equation}
Y_{i}^{G}=\tilde{Y}_{i+1}^{G}+\left( N_{i}^{G} -N_{i+1}^{G}       \right),
\end{equation}
where $N_{i}^{G} - N_{i+1}^{G}$ is a Gaussian random vector with covariance matrix $\boldsymbol{K}_{i} - \boldsymbol{K}_{i+1}$, and is independent of $\tilde{Y}_{i+1}^{G}$.
Therefore, the covariance matrix of
$$
\left(\begin{array}{r}
 \sqrt{{\gamma}/{(1-\gamma)}} \tilde{Y}_{i+1}^{G}\\
- \sqrt{{(1-\gamma)}/{\gamma}}    {{Y}}_{i}^{G}
\end{array}\right)
$$
is given by
\begin{align}
\tilde{\boldsymbol{\Sigma}}_{i} &\triangleq \begin{pmatrix}
\frac{\gamma}{{1-\gamma}}\left( \boldsymbol{\Delta}_{i}+ \boldsymbol{K}_{i}\right) & -\left( \boldsymbol{\Delta}_{i}+ \boldsymbol{K}_{i}\right)\\
- \left( \boldsymbol{\Delta}_{i}+ \boldsymbol{K}_{i}\right) & \frac{1-\gamma}{\gamma}\left( \boldsymbol{\Delta}_{i} +\boldsymbol{K}_{i+1} \right)
\end{pmatrix}. \label{eq:covm2}
\end{align}
It can be verified that
\begin{equation}\label{eq:covi2}
\tilde{\boldsymbol{\Sigma}}_{i}^{-1} =
 \begin{pmatrix}
\frac{1-\gamma}{\gamma}\left( \left(\boldsymbol{\Delta}_{i} + \boldsymbol{K}_{i+1}\right)^{-1} + \left(  \boldsymbol{K}_{i} - \boldsymbol{K}_{i+1}\right)^{-1}     \right) & \left(  \boldsymbol{K}_{i} - \boldsymbol{K}_{i+1}\right)^{-1} \\
\left(  \boldsymbol{K}_{i} - \boldsymbol{K}_{i+1}\right)^{-1}  & \frac{\gamma}{1-\gamma}  \left(  \boldsymbol{K}_{i} - \boldsymbol{K}_{i+1}\right)^{-1}
\end{pmatrix}
\end{equation}
and
\begin{equation}\label{eq:covd2}
\nabla_{\gamma}{\tilde{\boldsymbol{\Sigma}}_{i} } =  \begin{pmatrix}
\frac{1}{(1-\gamma)^{2}}\left( \boldsymbol{\Delta}_{i}+\boldsymbol{K}_{i+1} \right) & \boldsymbol{0}\\
\boldsymbol{0} & -\frac{1}{\gamma^{2}} (  \boldsymbol{\Delta}_{i} + \boldsymbol{K}_{i}   )
\end{pmatrix}.
\end{equation}
Combining \eqref{eq:covi2} and \eqref{eq:covd2} gives
\begin{equation}\label{eq:covm12}
\tr \left\{ \left(\nabla_{\gamma}{\tilde{\boldsymbol{\Sigma}}_{i} }\right) \tilde{\boldsymbol{\Sigma}}_{i}^{-1}  \right\}=0,
\end{equation}
\begin{equation}\label{eq:covm22}
\tilde{\boldsymbol{\Sigma}}_{i}^{-1}\left( \nabla_{\gamma}{\tilde{\boldsymbol{\Sigma}}_{i} } \right) \tilde{\boldsymbol{\Sigma}}_{i}^{-1} = \begin{pmatrix}
-\frac{1}{\gamma^{2}}\left( \left(\boldsymbol{\Delta}_{i} + \boldsymbol{K}_{i+1}\right)^{-1} + \left(  \boldsymbol{K}_{i} - \boldsymbol{K}_{i+1}\right)^{-1}     \right)& \boldsymbol{0}\\
\boldsymbol{0} & \frac{1}{(1-\gamma)^{2}}  \left(  \boldsymbol{K}_{i} - \boldsymbol{K}_{i+1}\right)^{-1}
\end{pmatrix}.
\end{equation}

By invoking the chain rule of matrix calculus and Lemma \ref{de} in Appendix \ref{app_lea2}, we have
\begin{align}
&\frac{d}{d\gamma}h\left(\tilde{Y}_{i+1, \gamma}, {Y}^{*}_{i, \gamma} | W_{j}, j \in [1:i]\right) \nonumber \\
&= \frac{d}{d\gamma} \left\{ h\left(        Y_{i+1} + \sqrt{\frac{\gamma}{1-\gamma}} \tilde{Y}_{i+1}^{G}, Y_{i} - \sqrt{\frac{1-\gamma}{\gamma}}    {{Y}}_{i}^{G}   \Big| W_{j}, j \in [1:i]                        \right)+\frac{n}{2} \log \gamma + \frac{n}{2} \log (1-\gamma) \right\} \\
&= \frac{1}{2} \tr \left\{\left( \nabla_{\gamma}\tilde{\boldsymbol{\Sigma}}_{i} \right)  J \left(
\begin{pmatrix}
\sqrt{\frac{1}{1-\gamma}}{\tilde{Y}^{T}_{i+1, \gamma}} & \sqrt{\frac{1}{\gamma}}{{Y}^{*}_{i, \gamma}}^{T}
\end{pmatrix}^{T}
\Big| W_{j}, j \in [1:i]                        \right)            \right\} + \frac{n}{2} \left(  \frac{1}{\gamma} - \frac{1}{1-\gamma} \right)\label{eq:covm5}
\end{align}
It can be verified that
\begin{align}
&  \tr \left\{\left( \nabla_{\gamma}\tilde{\boldsymbol{\Sigma}}_{i} \right)  J \left(
\begin{pmatrix}
\sqrt{\frac{1}{1-\gamma}}{\tilde{Y}^{T}_{i+1, \gamma}} & \sqrt{\frac{1}{\gamma}}{{Y}^{*}_{i, \gamma}}^{T}
\end{pmatrix}^{T}
\Big| W_{j}, j \in [1:i]                        \right)            \right\}\nonumber \\
& = \tr  \Big\{ \left(\nabla_{\gamma}{\tilde{\boldsymbol{\Sigma}}_{i} }\right) \tilde{\boldsymbol{\Sigma}}_{i}^{-1}  -    \tilde{\boldsymbol{\Sigma}}_{i}^{-1} \left( \nabla_{\gamma}\tilde{{\boldsymbol{\Sigma}}}_{i} \right) \tilde{\boldsymbol{\Sigma}}_{i}^{-1}\nonumber \\
&\qquad \qquad \cov \left(
 \begin{pmatrix}
Y_{i+1}^{T} & Y_{i}^{T}
\end{pmatrix}^{T}
 \Big|   Y_{i+1} + \sqrt{\frac{\gamma}{1-\gamma}} \tilde{Y}_{i+1}^{G}, Y_{i} - \sqrt{\frac{1-\gamma}{\gamma}}    {{Y}}_{i}^{G}, W_{j}, j \in [1:i]     \right)       \Big\} \label{eq:compli2}\\
&= \tr \Big\{           \begin{pmatrix}
-\frac{1}{\gamma^{2}}\left( \left(\boldsymbol{\Delta}_{i} + \boldsymbol{K}_{i+1}\right)^{-1} + \left(  \boldsymbol{K}_{i} - \boldsymbol{K}_{i+1}\right)^{-1}     \right)& \boldsymbol{0}\\
\boldsymbol{0} & \frac{1}{(1-\gamma)^{2}}  \left(  \boldsymbol{K}_{i} - \boldsymbol{K}_{i+1}\right)^{-1}
\end{pmatrix} \nonumber\\
&
\qquad \qquad \cov \left(
 \begin{pmatrix}
Y_{i+1}^{T} & Y_{i}^{T}
\end{pmatrix}^{T}
 \Big|   Y_{i+1} + \sqrt{\frac{\gamma}{1-\gamma}} \tilde{Y}_{i+1}^{G}, Y_{i} - \sqrt{\frac{1-\gamma}{\gamma}}    {{Y}}_{i}^{G}, W_{j}, j \in [1:i]     \right)                   \Big\}, \label{eq:covm3}
\end{align}
where \eqref{eq:compli2} follows by Lemma \ref{lea1} in Appendix \ref{app_lea2}, and \eqref{eq:covm3} is due to \eqref{eq:covm12} and \eqref{eq:covm22}.
Notice that
\begin{align}
& \cov \left(
 \begin{pmatrix}
Y_{i+1}^{T} & Y_{i}^{T}
\end{pmatrix}^{T}
\Big|    Y_{i+1} + \sqrt{\frac{\gamma}{1-\gamma}} \tilde{Y}_{i+1}^{G}, Y_{i} - \sqrt{\frac{1-\gamma}{\gamma}}    {{Y}}_{i}^{G}   \right)   \nonumber \\
& =
\left(
\begin{pmatrix}
\boldsymbol{K}_{0}+\boldsymbol{K}_{i+1} & \boldsymbol{K}_{0}+\boldsymbol{K}_{i+1} \\
\boldsymbol{K}_{0}+\boldsymbol{K}_{i+1} & \boldsymbol{K}_{0}+ \boldsymbol{K}_{i}
\end{pmatrix}^{-1}
+ \tilde{\boldsymbol{\Sigma}}_{i}^{-1}
\right)^{-1} \\
& =
\left(
\begin{pmatrix}
\left(\boldsymbol{K}_{0}+\boldsymbol{K}_{i+1}\right)^{-1}+\left(\boldsymbol{K}_{i}-\boldsymbol{K}_{i+1}\right)^{-1} & -\left(\boldsymbol{K}_{i}-\boldsymbol{K}_{i+1}\right)^{-1} \\
-\left(\boldsymbol{K}_{i}-\boldsymbol{K}_{i+1}\right)^{-1} & \left(\boldsymbol{K}_{i}-\boldsymbol{K}_{i+1}\right)^{-1}
\end{pmatrix}\right.
\nonumber \\
& \quad \left.+
\begin{pmatrix}
\frac{1-\gamma}{\gamma}\left( \left(\boldsymbol{\Delta}_{i} + \boldsymbol{K}_{i+1}\right)^{-1} + \left(  \boldsymbol{K}_{i} - \boldsymbol{K}_{i+1}\right)^{-1}     \right) & \left(  \boldsymbol{K}_{i} - \boldsymbol{K}_{i+1}\right)^{-1} \\
\left(  \boldsymbol{K}_{i} - \boldsymbol{K}_{i+1}\right)^{-1}  & \frac{\gamma}{1-\gamma}  \left(  \boldsymbol{K}_{i} - \boldsymbol{K}_{i+1}\right)^{-1}
\end{pmatrix}
\right)^{-1} \\
& =
\begin{pmatrix}
\left(  \left(\boldsymbol{K}_{0}+ \boldsymbol{K}_{i+1}\right)^{-1}+  \frac{1-\gamma}{\gamma} \left(\boldsymbol{\Delta}_{i} + \boldsymbol{K}_{i+1}\right)^{-1} + \frac{1}{\gamma}\left(  \boldsymbol{K}_{i} - \boldsymbol{K}_{i+1}\right)^{-1}         \right)^{-1} & \boldsymbol{0}\\
\boldsymbol{0} & (1-\gamma)  \left(\boldsymbol{K}_{i}-\boldsymbol{K}_{i+1}\right)
\end{pmatrix}.\label{eq_152}
\end{align}
Thus, we have the Markov chain
\begin{equation}
 \left( W_{j}, j \in [1:i] \right) \rightarrow Y_{i+1} \rightarrow \left(  Y_{i+1} + \sqrt{\frac{\gamma}{1-\gamma}} \tilde{Y}_{i+1}^{G}, Y_{i} - \sqrt{\frac{1-\gamma}{\gamma}}    {{Y}}_{i}^{G}     \right) \rightarrow Y_{i}.
\end{equation}
As a consequence,
\begin{align}
& \cov \left(
 \begin{pmatrix}
 Y_{i+1}^{T} & Y_{i}^{T}
\end{pmatrix}^{T}
 \Big|    \tilde{Y}_{i+1, \gamma}, {Y}^{*}_{i, \gamma},  W_{j}, j \in [1:i]     \right)     \nonumber \\
& =
\begin{pmatrix}
\cov \left(   Y_{i+1} \Big | \tilde{Y}_{i+1, \gamma}, {Y}^{*}_{i, \gamma}, W_{j}, j \in [1:i]     \right) & \boldsymbol{0}\\
\boldsymbol{0} &  (1-\gamma)  (\boldsymbol{K}_{i}- \boldsymbol{K}_{i+1}) \label{eq:covm4}
\end{pmatrix}.
\end{align}
Combining \eqref{eq:covm5}, \eqref{eq:covm3} and \eqref{eq:covm4}, we obtain
\begin{align}
&\frac{d}{d\gamma}h\left(\tilde{Y}_{i+1, \gamma}, {Y}^{*}_{i, \gamma}\Big| W_{j}, j \in [1:i]\right) \nonumber \\
& = \frac{1}{2} \tr \left\{
 \begin{pmatrix}
-\frac{1}{\gamma^{2}}\left( \left(\boldsymbol{\Delta}_{i} + \boldsymbol{K}_{i+1}\right)^{-1} + \left(  \boldsymbol{K}_{i} - \boldsymbol{K}_{i+1}\right)^{-1}     \right)& \boldsymbol{0}\\
\boldsymbol{0} & \frac{1}{(1-\gamma)^{2}}  \left(  \boldsymbol{K}_{i} - \boldsymbol{K}_{i+1}\right)^{-1}
\end{pmatrix}\right. \nonumber\\
& \left.
\qquad \qquad
\begin{pmatrix}
\cov \left(   Y_{i+1} \Big | \tilde{Y}_{i+1, \gamma}, {Y}^{*}_{i, \gamma}, W_{j}, j \in [1:i]     \right) & \boldsymbol{0}\\
\boldsymbol{0} &  (1-\gamma)  (\boldsymbol{K}_{i}- \boldsymbol{K}_{i+1}) \label{eq:covmf}
\end{pmatrix}
\right\} + \frac{n}{2} \left(  \frac{1}{\gamma} - \frac{1}{1-\gamma} \right)   \\
& = -\frac{1}{2\gamma} \tr \left\{   \frac{1}{\gamma}\left( \left(\boldsymbol{\Delta}_{i} + \boldsymbol{K}_{i+1}\right)^{-1} + \left(  \boldsymbol{K}_{i} - \boldsymbol{K}_{i+1}\right)^{-1}     \right) \cov \left(   Y_{i+1} \Big | \tilde{Y}_{i+1, \gamma}, {Y}^{*}_{i, \gamma}, W_{j}, j \in [1:i]     \right) \ -\boldsymbol{I}   \right\}.\label{eq:tt2.2}
\end{align}

On the other hand, it follows by the theory of linear MMSE estimation that
\begin{align}
\sqrt{\gamma}\tilde{Y}_{i+1}^{G}
=& -\sqrt{\gamma (1-\gamma)} \left(    \left(\boldsymbol{\Delta}_{i}+\boldsymbol{K}_{i+1}\right)^{-1} + \left( 1-\gamma \right)\left(\boldsymbol{K}_{i}-\boldsymbol{K}_{i+1}\right)^{-1}   \right)^{-1}\left(\boldsymbol{K}_{i}-\boldsymbol{K}_{i+1}\right)^{-1} \nonumber \\
&\left( \sqrt{\gamma} N_{i} -\sqrt{\gamma} N_{i+1}-\sqrt{1-\gamma}Y_{i}^{G}      \right) + \sqrt{\gamma}\hat{Y}_{i+1}^{G},
\end{align}
where $\hat{Y}_{i+1,\gamma}$ is a Gaussian random vector with mean zero and covariance matrix $  \left(    \left(\boldsymbol{\Delta}_{i}+\boldsymbol{K}_{i+1}\right)^{-1} + \left( 1-\gamma \right)\left(\boldsymbol{K}_{i}-\boldsymbol{K}_{i+1}\right)^{-1}   \right)^{-1}$, and is independent of $\sqrt{\gamma} \left(N_{i}-N_{i+1}\right) -\sqrt{1-\gamma}Y_{i}^{G}  $.
Thus, we have
\begin{align}
&\tilde{Y}_{i+1}= \sqrt{1-\gamma}Y_{i+1} + \sqrt{\gamma}\tilde{Y}_{i+1}^{G} \nonumber \\
&=\sqrt{1-\gamma}Y_{i+1}-\sqrt{\gamma (1-\gamma)} \left(    \left(\boldsymbol{\Delta}_{i}+\boldsymbol{K}_{i+1}\right)^{-1} + \left( 1-\gamma \right)\left(\boldsymbol{K}_{i}-\boldsymbol{K}_{i+1}\right)^{-1}   \right)^{-1}\left(\boldsymbol{K}_{i}-\boldsymbol{K}_{i+1}\right)^{-1} \nonumber \\
&\quad \left( \sqrt{\gamma} N_{i} -\sqrt{\gamma} N_{i+1}-\sqrt{1-\gamma}Y_{i}^{G}      \right) + \sqrt{\gamma}\hat{Y}_{i+1}^{G}\\
&=\sqrt{1-\gamma}   \left(    \left(\boldsymbol{\Delta}_{i}+\boldsymbol{K}_{i+1}\right)^{-1} + \left( 1-\gamma \right)\left(\boldsymbol{K}_{i}-\boldsymbol{K}_{i+1}\right)^{-1}   \right)^{-1} \left(    \left(\boldsymbol{\Delta}_{i}+\boldsymbol{K}_{i+1}\right)^{-1} + \left(\boldsymbol{K}_{i}-\boldsymbol{K}_{i+1}\right)^{-1}   \right)Y_{i+1}\nonumber \\
&\quad + \sqrt{\gamma}\hat{Y}_{i+1}^{G}-\sqrt{\gamma (1-\gamma)} \left(    \left(\boldsymbol{\Delta}_{i}+\boldsymbol{K}_{i+1}\right)^{-1} + \left( 1-\gamma \right)\left(\boldsymbol{K}_{i}-\boldsymbol{K}_{i+1}\right)^{-1}   \right)^{-1}\left(\boldsymbol{K}_{i}-\boldsymbol{K}_{i+1}\right)^{-1}Y^{*}_{i,\gamma}.
\end{align}
The complementary Fisher information representation of $\cov \left(   Y_{i+1} \Big | \tilde{Y}_{i+1, \gamma}, {Y}^{*}_{i, \gamma}, W_{j}, j \in [1:i]     \right)$ can be thereby expressed as
\begin{align}
&\cov \left(   Y_{i+1} \Big | \tilde{Y}_{i+1, \gamma}, {Y}^{*}_{i, \gamma}, W_{j}, j \in [1:i]     \right) \nonumber \\
&=\frac{\gamma}{1-\gamma}\left(    \left(\boldsymbol{\Delta}_{i}+\boldsymbol{K}_{i+1}\right)^{-1} +\left(\boldsymbol{K}_{i}-\boldsymbol{K}_{i+1}\right)^{-1}   \right)^{-1}     \left(    \left(\boldsymbol{\Delta}_{i}+\boldsymbol{K}_{i+1}\right)^{-1} + \left( 1-\gamma \right)\left(\boldsymbol{K}_{i}-\boldsymbol{K}_{i+1}\right)^{-1} -\right. \nonumber \\
& \qquad \left. \gamma J\left(    \tilde{Y}_{i+1, \gamma} \Big |  {Y}^{*}_{i, \gamma},  W_{j}, j \in [1:i]  \right)          \right) \left(    \left(\boldsymbol{\Delta}_{i}+\boldsymbol{K}_{i+1}\right)^{-1} +\left(\boldsymbol{K}_{i}-\boldsymbol{K}_{i+1}\right)^{-1}   \right)^{-1}. \label{eq:tt2}
\end{align}
Equivalently, we can write
\begin{align}
&\left(    \left(\boldsymbol{\Delta}_{i}+\boldsymbol{K}_{i+1}\right)^{-1} +\left(\boldsymbol{K}_{i}-\boldsymbol{K}_{i+1}\right)^{-1}   \right)
\left(  \frac{1}{\gamma}  \cov \left(   Y_{i+1} \Big | \tilde{Y}_{i+1, \gamma}, {Y}^{*}_{i, \gamma}, W_{j}, j \in [1:i]     \right)    \right. \nonumber \\
& \quad \left.   -  \left(    \left(\boldsymbol{\Delta}_{i}+\boldsymbol{K}_{i+1}\right)^{-1} +\left(\boldsymbol{K}_{i}-\boldsymbol{K}_{i+1}\right)^{-1}   \right)^{-1}  \right)
\left(    \left(\boldsymbol{\Delta}_{i}+\boldsymbol{K}_{i+1}\right)^{-1} +\left(\boldsymbol{K}_{i}-\boldsymbol{K}_{i+1}\right)^{-1}   \right) \nonumber\\
&=\frac{\gamma}{1-\gamma}\left(\boldsymbol{\Delta}_{i}+\boldsymbol{K}_{i+1}\right)^{-1}-\frac{\gamma}{1-\gamma}J\left(    \tilde{Y}_{i+1, \gamma} \Big |  {Y}^{*}_{i, \gamma},  W_{j}, j \in [1:i]  \right). \label{eq:tt2.1}
\end{align}
Substituting \eqref{eq:tt2.1} into \eqref{eq:tt2.2} gives
\begin{align}
&\frac{d}{d\gamma}h\left(\tilde{Y}_{i+1, \gamma} , {Y}^{*}_{i, \gamma}\Big| W_{j}, j \in [1:i]\right) \nonumber \\
&=\frac{1}{2(1-\gamma)} \tr \left\{   \left(    \left(\boldsymbol{\Delta}_{i}+\boldsymbol{K}_{i+1}\right)^{-1} +\left(\boldsymbol{K}_{i}-\boldsymbol{K}_{i+1}\right)^{-1}   \right)^{-1} \left(  J\left(  \tilde{Y}_{i+1, \gamma} \Big |  {Y}^{*}_{i, \gamma},  W_{j}, j \in [1:i]   \right) \ -  \left(\boldsymbol{\Delta}_{i}+\boldsymbol{K}_{i+1}\right)^{-1} \right)  \right\}.\label{eq:tt2.3}
\end{align}
Furthermore, it follows by the Woodbury matrix inversion lemma that
\begin{align}
&\left( \left(\boldsymbol{\Delta}_{i} + \boldsymbol{K}_{i+1}\right)^{-1} + \left(  \boldsymbol{K}_{i} - \boldsymbol{K}_{i+1}\right)^{-1} \right)^{-1} \nonumber \\
&=\boldsymbol{K}_{i+1}\left(\boldsymbol{K}_{i+1}-\boldsymbol{K}_{i+1}\left(  \boldsymbol{K}_{i+1} - \boldsymbol{K}_{i}\right)^{-1}\boldsymbol{K}_{i+1} - \boldsymbol{K}_{i+1} + \boldsymbol{K}_{i+1}\left(\boldsymbol{\Delta}_{i} + \boldsymbol{K}_{i+1}\right)^{-1}\boldsymbol{K}_{i+1}\right)^{-1}\boldsymbol{K}_{i+1}\\
&=\boldsymbol{K}_{i+1} \left(    \left( \boldsymbol{K}_{i+1}^{-1} - \boldsymbol{K}_{i}^{-1}          \right)^{-1}     -          \left( \boldsymbol{\Delta}_{i}^{-1}+\boldsymbol{K}_{i+1}^{-1}           \right)^{-1}               \right)^{-1}\boldsymbol{K}_{i+1} \\
&=\boldsymbol{K}_{i+1}\left( \boldsymbol{\Delta}_{i}^{-1}+\boldsymbol{K}_{i+1}^{-1}           \right) \left(  \left( \boldsymbol{\Delta}_{i}^{-1}+\boldsymbol{K}_{i}^{-1}           \right)^{-1} - \left( \boldsymbol{\Delta}_{i}^{-1}+\boldsymbol{K}_{i+1}^{-1}           \right)^{-1}   \right)\left( \boldsymbol{\Delta}_{i}^{-1}+\boldsymbol{K}_{i+1}^{-1}           \right)\boldsymbol{K}_{i+1}.
\end{align}
So we can rewrite \eqref{eq:tt2.3}  as
\begin{align}
&\frac{d}{d\gamma}h\left(\tilde{Y}_{i+1, \gamma} , {Y}^{*}_{i, \gamma}\Big| W_{j}, j \in [1:i]\right) \nonumber \\
&=\frac{1}{2(1-\gamma)} \tr \Big\{   \left(  \left( \boldsymbol{\Delta}_{i}^{-1}+\boldsymbol{K}_{i}^{-1}           \right)^{-1} - \left( \boldsymbol{\Delta}_{i}^{-1}+\boldsymbol{K}_{i+1}^{-1}           \right)^{-1}   \right)\Big(\left( \boldsymbol{\Delta}_{i}^{-1}+\boldsymbol{K}_{i+1}^{-1}           \right)\boldsymbol{K}_{i+1}\nonumber\\
& \qquad  J\left(  \tilde{Y}_{i+1, \gamma} \Big |  {Y}^{*}_{i, \gamma},  W_{j}, j \in [1:i]   \right)\boldsymbol{K}_{i+1} \left( \boldsymbol{\Delta}_{i}^{-1}+\boldsymbol{K}_{i+1}^{-1}           \right)\ - \boldsymbol{\Delta}_{i}^{-1} \left(\boldsymbol{\Delta}_{i}+\boldsymbol{K}_{i+1}\right) \boldsymbol{\Delta}_{i}^{-1}\Big) \Big\}.
\end{align}

\bibliographystyle{IEEEtran}
\bibliography{ref}

\begin{thebibliography}{10}
\providecommand{\url}[1]{#1}
\csname url@samestyle\endcsname
\providecommand{\newblock}{\relax}
\providecommand{\bibinfo}[2]{#2}
\providecommand{\BIBentrySTDinterwordspacing}{\spaceskip=0pt\relax}
\providecommand{\BIBentryALTinterwordstretchfactor}{4}
\providecommand{\BIBentryALTinterwordspacing}{\spaceskip=\fontdimen2\font plus
\BIBentryALTinterwordstretchfactor\fontdimen3\font minus
  \fontdimen4\font\relax}
\providecommand{\BIBforeignlanguage}[2]{{%
\expandafter\ifx\csname l@#1\endcsname\relax
\typeout{** WARNING: IEEEtran.bst: No hyphenation pattern has been}%
\typeout{** loaded for the language `#1'. Using the pattern for}%
\typeout{** the default language instead.}%
\else
\language=\csname l@#1\endcsname
\fi
#2}}
\providecommand{\BIBdecl}{\relax}
\BIBdecl

\bibitem{SW73}
D.~Slepian and J.~Wolf, ``Noiseless coding of correlated information sources,''
  \emph{{IEEE} Trans. Inf. Theory}, vol.~19, no.~4, pp. 471--480, Jul. 1973.

\bibitem{WZ76}
A.~Wyner and J.~Ziv, ``The rate-distortion function for source coding with side
  information at the decoder,'' \emph{{IEEE} Trans. Inf. Theory}, vol.~22,
  no.~1, pp. 1--10, Jan. 1976.

\bibitem{SM04}
Y.~Steinberg and N.~Merhav, ``On successive refinement for the {W}yner-{Z}iv
  problem,'' \emph{{IEEE} Trans. Inf. Theory}, vol.~50, no.~8, pp. 1636--1654,
  Aug. 2004.

\bibitem{TD07}
C.~Tian and S.~N. Diggavi, ``On multistage successive refinement for
  {W}yner-{Z}iv source coding with degraded side informations,'' \emph{{IEEE}
  Trans. Inf. Theory}, vol.~53, no.~8, pp. 2946--2960, Aug. 2007.

\bibitem{Bergmans74}
P.~Bergmans, ``A simple converse for broadcast channels with additive white
  {G}aussian noise (corresp.),'' \emph{{IEEE} Trans. Inf. Theory}, vol.~20,
  no.~2, pp. 279--280, May 1974.

\bibitem{WLSSV09}
H.~{Weingarten}, T.~{Liu}, S.~{Shamai}, Y.~{Steinberg}, and P.~{Viswanath},
  ``The capacity region of the degraded multiple-input multiple-output compound
  broadcast channel,'' \emph{{IEEE} Trans. Inf. Theory}, vol.~55, no.~11, pp.
  5011--5023, Nov. 2009.

\bibitem{LLL09}
H.~D. {Ly}, T.~{Liu}, and Y.~{Liang}, ``Multiple-input multiple-output
  {G}aussian broadcast channels with common and confidential messages,''
  \emph{{IEEE} Trans. Inf. Theory}, vol.~56, no.~11, pp. 5477--5487, Nov. 2010.

\bibitem{EU12-1}
E.~{Ekrem} and S.~{Ulukus}, ``Capacity region of {G}aussian {MIMO} broadcast
  channels with common and confidential messages,'' \emph{{IEEE} Trans. Inf.
  Theory}, vol.~58, no.~9, pp. 5669--5680, Sep. 2012.

\bibitem{EU12-2}
------, ``Capacity-equivocation region of the {G}aussian {MIMO} wiretap
  channel,'' \emph{{IEEE} Trans. Inf. Theory}, vol.~58, no.~9, pp. 5699--5710,
  Sep. 2012.

\bibitem{LLPS13}
R.~{Liu}, T.~{Liu}, H.~V. {Poor}, and S.~{Shamai}, ``New results on
  multiple-input multiple-output broadcast channels with confidential
  messages,'' \emph{{IEEE} Trans. Inf. Theory}, vol.~59, no.~3, pp. 1346--1359,
  Mar. 2013.

\bibitem{EU13}
E.~Ekrem and S.~Ulukus, ``Secure lossy transmission of vector {G}aussian
  sources,'' \emph{{IEEE} Trans. Inf. Theory}, vol.~59, no.~9, pp. 5466--5487,
  Sep. 2013.

\bibitem{CL14-1}
H.~{Chong} and Y.~{Liang}, ``The capacity region of the class of three-receiver
  {G}aussian {MIMO} multilevel broadcast channels with two-degraded message
  sets,'' \emph{{IEEE} Trans. Inf. Theory}, vol.~60, no.~1, pp. 42--53, Jan.
  2014.

\bibitem{CL14-2}
------, ``An extremal inequality and the capacity region of the degraded
  compound {G}aussian {MIMO} broadcast channel with multiple users,''
  \emph{{IEEE} Trans. Inf. Theory}, vol.~60, no.~10, pp. 6131--6143, Oct. 2014.

\bibitem{KL14}
A.~{Khisti} and T.~{Liu}, ``Private broadcasting over independent parallel
  channels,'' \emph{{IEEE} Trans. Inf. Theory}, vol.~60, no.~9, pp. 5173--5187,
  Sep. 2014.

\bibitem{MK09}
A.~S. Motahari and A.~K. Khandani, ``Capacity bounds for the {G}aussian
  interference channel,'' \emph{{IEEE} Trans. Inf. Theory}, vol.~55, no.~2, pp.
  620--643, Feb. 2009.

\bibitem{SKC09}
X.~Shang, G.~Kramer, and B.~Chen, ``A new outer bound and the
  noisy-interference sum-rate capacity for {G}aussian interference channels,''
  \emph{{IEEE} Trans. Inf. Theory}, vol.~55, no.~2, pp. 689--699, Feb. 2009.

\bibitem{AV09}
V.~S. Annapureddy and V.~V. Veeravalli, ``{G}aussian interference networks:
  {S}um capacity in the low interference regime and new outer bounds on the
  capacity region,'' \emph{{IEEE} Trans. Inf. Theory}, vol.~55, no.~7, pp.
  3032--3050, Jul. 2009.

\bibitem{Oohama05}
Y.~Oohama, ``Rate-distortion theory for {G}aussian multiterminal source coding
  systems with several side informations at the decoder,'' \emph{{IEEE} Trans.
  Inf. Theory}, vol.~51, no.~7, pp. 2577--2593, Jul. 2005.

\bibitem{WCW10}
J.~Wang, J.~Chen, and X.~Wu, ``On the sum rate of {G}aussian multiterminal
  source coding: {N}ew proofs and results,'' \emph{{IEEE} Trans. Inf. Theory},
  vol.~56, no.~8, pp. 3946--3960, Aug. 2010.

\bibitem{XW13}
Y.~Xu and Q.~Wang, ``A perturbation proof of the vector {G}aussian one-help-one
  problem,'' in \emph{Proc. {IEEE} Int. Symp. Inf. Theory}, Istanbul, Turkey,
  Jul. 2013.

\bibitem{WC13}
J.~Wang and J.~Chen, ``Vector {G}aussian two-terminal source coding,''
  \emph{{IEEE} Trans. Inf. Theory}, vol.~59, no.~6, pp. 3693--3708, Jun. 2013.

\bibitem{WC14}
------, ``Vector {G}aussian multiterminal source coding,'' \emph{{IEEE} Trans.
  Inf. Theory}, vol.~60, no.~9, pp. 5533--5552, Sep. 2014.

\bibitem{WO11}
S.~Watanabe and Y.~Oohama, ``Secret key agreement from vector {G}aussian
  sources by rate limited public communication,'' \emph{{IEEE} Trans. Inf.
  Forensics Security}, vol.~6, no.~3, pp. 541--550, Sep. 2011.

\bibitem{Ozarow80}
L.~{Ozarow}, ``On a source coding problem with two channels and three
  receivers,'' \emph{Bell Syst. Tech. J.}, vol.~59, no.~10, pp. 1909--1921,
  Dec. 1980.

\bibitem{WV07}
H.~{Wang} and P.~{Viswanath}, ``Vector {G}aussian multiple description with
  individual and central receivers,'' \emph{{IEEE} Trans. Inf. Theory},
  vol.~53, no.~6, pp. 2133--2153, Jun. 2007.

\bibitem{C09}
J.~{Chen}, ``Rate region of {G}aussian multiple description coding with
  individual and central distortion constraints,'' \emph{{IEEE} Trans. Inf.
  Theory}, vol.~55, no.~9, pp. 3991--4005, Sep. 2009.

\bibitem{XCW17}
Y.~{Xu}, J.~{Chen}, and Q.~{Wang}, ``The sum rate of vector {G}aussian multiple
  description coding with tree-structured covariance distortion constraints,''
  \emph{{IEEE} Trans. Inf. Theory}, vol.~63, no.~10, pp. 6547--6560, Oct. 2017.

\bibitem{SCWL13}
L.~Song, J.~Chen, J.~Wang, and T.~Liu, ``{G}aussian robust sequential and
  predictive coding,'' \emph{{IEEE} Trans. Inf. Theory}, vol.~59, no.~6, pp.
  3635--3652, Jun. 2013.

\bibitem{SCT15}
L.~{Song}, J.~{Chen}, and C.~{Tian}, ``Broadcasting correlated vector
  {G}aussians,'' \emph{{IEEE} Trans. Inf. Theory}, vol.~61, no.~5, pp.
  2465--2477, May 2015.

\bibitem{GN14}
Y.~{Geng} and C.~{Nair}, ``The capacity region of the two-receiver {G}aussian
  vector broadcast channel with private and common messages,'' \emph{{IEEE}
  Trans. Inf. Theory}, vol.~60, no.~4, pp. 2087--2104, Apr. 2014.

\bibitem{WC19}
\BIBentryALTinterwordspacing
J.~Wang and J.~Chen, ``A monotone path proof of an extremal result for long
  {M}arkov chains,'' \emph{Entropy}, vol.~21, no.~3, 2019. [Online]. Available:
  \url{http://www.mdpi.com/1099-4300/21/3/276}
\BIBentrySTDinterwordspacing

\bibitem{C18}
T.~A. {Courtade}, ``A strong entropy power inequality,'' \emph{{IEEE} Trans.
  Inf. Theory}, vol.~64, no.~4, pp. 2173--2192, Apr. 2018.

\bibitem{WSS06}
H.~Weingarten, Y.~Steinberg, and S.~S. Shamai, ``The capacity region of the
  {G}aussian multiple-input multiple-output broadcast channel,'' \emph{{IEEE}
  Trans. Inf. Theory}, vol.~52, no.~9, pp. 3936--3964, Sep. 2006.

\bibitem{DCT91}
A.~Dembo, T.~Cover, and J.~Thomas, ``Information theoretic inequalities,''
  \emph{{IEEE} Trans. Inf. Theory}, vol.~37, no.~6, pp. 1501 --1518, Nov. 1991.

\bibitem{GSV05}
D.~Guo, S.~Shamai, and S.~Verd\'{u}, ``Mutual information and minimum
  mean-square error in {G}aussian channels,'' \emph{{IEEE} Trans. Inf. Theory},
  vol.~51, no.~4, pp. 1261--1282, Apr. 2005.

\bibitem{book}
D.~P. Bertsekas, A.~Nedi{\'c}, and A.~E. Ozdaglar, \emph{Convex Analysis and
  Optimization}.\hskip 1em plus 0.5em minus 0.4em\relax Athena Scientific
  Belmont, 2003.

\bibitem{UW17}
S.~Unal and A.~B. Wagner, ``Vector {G}aussian rate-distortion with variable
  side information,'' \emph{{IEEE} Trans. Inf. Theory}, vol.~63, no.~8, pp.
  5162--5178, Aug. 2017.

\bibitem{PV06}
D.~Palomar and S.~Verd\'{u}, ``Gradient of mutual information in linear vector
  {G}aussian channels,'' \emph{{IEEE} Trans. Inf. Theory}, vol.~52, no.~1, pp.
  141--154, 2006.

\bibitem{LV07}
T.~Liu and P.~Viswanath, ``An extremal inequality motivated by multiterminal
  information-theoretic problems,'' \emph{{IEEE} Trans. Inf. Theory}, vol.~53,
  no.~5, pp. 1839 --1851, May 2007.

\bibitem{Z98}
R.~Zamir, ``A proof of the {F}isher information inequality via a data
  processing argument,'' \emph{{IEEE} Trans. Inf. Theory}, vol.~44, no.~3, pp.
  1246 --1250, May 1998.

\bibitem{R11}
O.~Rioul, ``Information theoretic proofs of entropy power inequalities,''
  \emph{{IEEE} Trans. Inf. Theory}, vol.~57, no.~1, pp. 33--55, 2011.

\end{thebibliography}

\end{document}